# ReaxKit: A Modular Python Toolkit for Preparing, Parsing, and Analyzing ReaxFF Molecular Dynamics Simulations

Ali Mohammadi Dinani[1], Alireza Sepehrinezhad[2], Anirban Phukan[3], Asma Ul Hosna[4], Jupjeet Dhingra[4], Mozhdeh Mirakhory[4], Seyed Mahmoud Mortazavi[4], Yun Kyung Shin[4], Swarit Dwivedi[5], Adri C.T. van Duin[1,2,3,4*]

[1] Department of Chemical Engineering, The Pennsylvania State University, University Park, PA 16802, United States

[2] Department of Engineering Science and Mechanics, The Pennsylvania State University, University Park, Pennsylvania 16802, United States

[3] Department of Chemistry, Eberly College of Science, The Pennsylvania State University, University Park Pennsylvania 16802, United States

[4] Department of Mechanical Engineering, The Pennsylvania State University, University Park Pennsylvania 16802, United States

[5] Department of Materials Science and Engineering, Monash University, Clayton 3800, Australia

* Corresponding author email: acv13@psu.edu

Abstract

Empirical reactive force field (RFF) molecular dynamics enables atomistic simulation of bond breaking, bond formation, charge redistribution, and structural evolution in chemically complex systems. The ReaxFF method is arguably the most popular and transferable of the currently available RFF methods. However, routine use of ReaxFF often requires substantial manual effort to prepare inputs, interpret engine-specific outputs, organize simulation artifacts, and develop custom analysis scripts, limiting reproducibility and scalability. Here, we present ReaxKit, a modular Python toolkit for preparing, parsing, analyzing, and managing ReaxFF molecular dynamics simulations. ReaxKit uses a separation-of-concerns architecture that distinguishes engine-specific input/output handling, canonical domain data models, scientific analysis, workflow orchestration, presentation, storage, and graphical interaction. Engine adapters convert outputs from supported simulation environments into typed, engine-independent data structures, allowing analysis modules to operate independently of native file formats. User requests are executed through consistent command-line, functional Python, and browser-based graphical interfaces, while a dedicated workspace preserves raw data, normalized datasets, analysis settings, results, logs, caches, and provenance information. Representative applications demonstrate the breadth of the toolkit, including automated generation of elastic and equation-of-state training data from Materials Project structures and mechanical properties, characterization of active sites and local structural environments, and execution of reproducible multi-parameter, multi-replicate simulation campaigns through YAML-defined study workflows. These capabilities show that ReaxKit supports multiple stages of the ReaxFF workflow, from input generation and simulation organization to scientific analysis, aggregation, visualization, and reporting. By reducing repetitive file handling and one-off scripting, ReaxKit provides an extensible foundation for reproducible, high-throughput, and potentially autonomous reactive molecular dynamics studies.

# 1. Introduction

Reactive force fields, particularly ReaxFF, are widely used to model chemically reactive systems at length and time scales inaccessible to quantum mechanical methods (Mao et al., 2023; Senftle et al., 2016). Since its introduction by van Duin and co-workers (Van Duin et al., 2001), who introduced a code that is currently referred to as the ‘Standalone ReaxFF code’, ReaxFF has provided a practical framework for reactive molecular dynamics simulations in which bond breaking, bond formation, charge redistribution, and structural evolution can be treated within a classical molecular dynamics framework. This has been accomplished through the ReaxFF bond-order formalism, where distance-dependent bond orders are computed and used within specific energy terms. As a result, ReaxFF is particularly useful for systems where chemical reactions, interfaces, defects formation and migration, and non-equilibrium processes play a central role.

Due to its capabilities especially in predicting reaction activation barriers, ReaxFF has been applied to a broad range of chemically complex systems, including catalysis (Shin et al., 2015), combustion (Chenoweth et al., 2008; Van Duin et al., 2001), corrosion (Verners & Van Duin, 2015), oxidation (Zou et al., 2015), materials growth (Momeni et al., 2022), polymers (Kowalik et al., 2019), batteries (Foss et al., 2025), nanoparticles (Hong & Van Duin, 2015), interface chemistry (Kulkarni et al., 2012), and ferroelectric materials with memery applications (Dryzhakov et al., 2026; Y. Liu et al., 2025) over the past two decades. In parallel with its growing scientific use, ReaxFF has been integrated into major simulation environments, including open-source platforms such as LAMMPS (Plimpton, 1995; Thompson et al., 2022) and commercial software packages such as the Amsterdam Modeling Suite (Baerends et al., 2025) (AMS) and BIOVIA Materials Studio through the GULP module (Gale, 1997; Gale & Rohl, 2003).

Despite this broad adoption, practical use of ReaxFF simulations still requires substantial effort in preparing, managing, and interpreting engine-specific input and output files. This challenge is especially significant when using the Standalone ReaxFF program (Van Duin et al., 2001), where simulations rely on multiple files with specific roles and formats. These files are often organized as simple text files with no column headers or variable definitions, and there is no developed package for managing them. As a results, users need to spend a good portion of their time going through the available manuals to learn which file contains which physical quantity, how different files are connected, and how to read their data and transform them into insights. Similar challenges also arise in other ReaxFF engines, such as LAMMPS or AMS, as each of them uses its own conventions, file formats, and schema for inputs and outputs.

Meanwhile, for computational chemists and materials scientists, the central objective is not to manually manage simulation files, organize data, and develop custom scripts but to efficiently run reactive simulations, extract chemically meaningful information, and compare results across systems, force fields, and simulation conditions. The lack of a user friendly, reusable, and engine-independent analysis infrastructure hence can increase the activation barrier for ReaxFF-related projects, limit its routine use, reduce reproducibility, and make high-throughput ReaxFF studies challenging. These problems in the long-term limit the wide adoption of ReaxFF especially by external groups and/or users with limited programming skills. Although specialized tools such as ChemTraYzer (for identifying and evaluating elementary reactions from ReaxFF trajectories) (Döntgen et al., 2015), ParAMS (for ReaxFF force field training) (Komissarov et al., 2021), and VARxMD (a tool for chemical reaction analysis and trajectories visualization) (J. Liu et al., 2014) have been developed, the need for an open-source modular, reusable, and engine-independent toolkit focused on the recurring file-processing and analysis tasks for ReaxFF simulations still remains.

Here, we introduce **ReaxKit**, a modular Python toolkit designed to bridge this gap by providing reusable tools for preparing, parsing, analyzing, and comparing ReaxFF molecular dynamics simulations across different different engines. Rather than replacing existing simulation engines or specialized tools, ReaxKit complements them by focusing on the recurring pre- and post-processing tasks that appear in routine and high-throughput ReaxFF studies: converting complex engine-specific outputs into analysis-ready data structures, extracting chemically meaningful quantities, and enabling reproducible comparison across simulations settings. In addition to post-processing capabilities, ReaxKit supports the generation of ReaxFF input files from user-defined settings, enabling more systematic and reproducible simulation setup. As will be demonstrated in next sections, this functionality allows ReaxKit to support end-to-end workflows in which input files are generated, simulations are executed, outputs are analyzed, and results are returned in a structured form. Apart from these, ReaxKit is accessible through multiple Application Programming Interfaces (APIs), including command-line workflows, a graphical user interface (GUI), and functional Python APIs, allowing users with different levels of programming experience to incorporate its tools into their routine ReaxFF use. With these features, ReaxKit serves as a lightweight and engine-independent workflow layer that reduces manual file handling, improves reproducibility and scalablility, and connects ReaxFF simulation outputs to both automated materials-research pipelines and specialized downstream analysis tools such as OVITO (Stukowski, 2010).

# 2. Methods

The following sections describe the main components of ReaxKit's layered software architecture, starting from engine adapters and analysis modules, then moving to workflow execution, utility functions, presentation and storage layers, graphical interfaces, and supporting package components. Finally, representative ReaxKit capabilities are presented to illustrate how these architectural elements work together in practical ReaxFF simulation workflows.

## 2.1. ReaxKit architecture overview

Main ReaxKit modules are grouped by role as shown in **Figure 1**, where they are organized into Engine, Analysis, Workflows, Utils, Presentation, and WebUI layers by functionality. It is worth mentioning that throughout the ReaxKit package, the word *ReaxFF* refers to *Standalone ReaxFF*, which is also reflected in this figure.

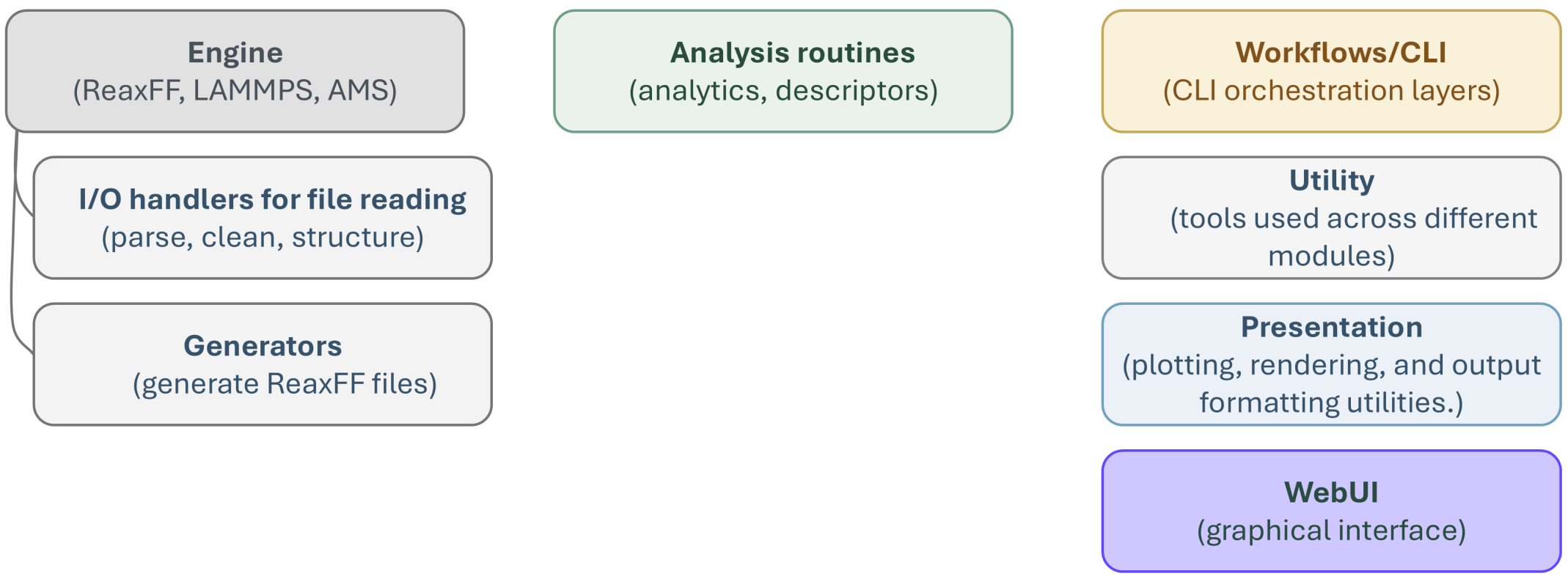

**Figure 1: Main module groups in ReaxKit.** ReaxKit is organized into engine-specific file handling and generation modules, reusable analysis routines, workflow/CLI orchestration layers, shared utilities, presentation tools, and a WebUI. This layered structure separates engine I/O from analysis and user interfaces, enabling ReaxKit to support different engines, analyzers, and presenters through a consistent and extensible software architecture.

The Engine layer contains file readers, writers, adapters, and generators for supported simulation engines, while the Analysis layer defines task-oriented computations that operate on normalized data structures. The Workflow layer connects user requests from command-line interface (CLI) to different commands, and the Presentation layer handles output formatting, plotting, and export. Utils are a set of utility logic shared between different modules, and the graphical user interface provides an accessible interaction layer for ReaxKit, allowing users to select input files, execute common analysis routines, and inspect generated outputs without requiring direct use of CLI or Python scripting. This organization reflects a separation-of-concerns design, in which engine-specific file handling and input generation, scientific analysis, workflow orchestration, shared utilities, presentation/export logic, and graphical interaction are implemented as distinct but interoperable, independently extensible layers.

Here are some sample files in each module:

- **Engine**: *xmolout_handler.py* is an Engine file because it parses a ReaxFF output (i.e., *xmolout* file) containing atomic trajectories.
- **Analysis**: *msd.py* is an Analysis file because it defines analysis classes and functions for computing mean square displacement (MSD) from trajectory data.
- **Workflows**: *trajectory_workflow.py* is a Workflow file because it handles CLI arguments and dispatches execution related to trajectory file handling (like writing or formatting a *xmolout* file) or trajectory-based analyses (like calculating MSD).
- **Utils**: *equation_of_states.py* is a Utils file because it provides reusable EOS functions used by training set generators, force field optimization analyzers, etc.
- **Presentation**: *scatter3d.py* is a Presentation file because it is responsible for a 3D plot rendering.
- **WebUI**: *execution_callbacks.py* is a WebUI file because it wires interactive UI callbacks to analysis execution actions.

## 2.2. Engine adapters with normalizers, file handlers, generators

ReaxKit manages engine-specific files through a unified adapter architecture implemented in *src/reaxkit/engine/base.py*, whose components are shown in the Figure S1. The generic *EngineAdapter* interface is the central abstraction and defines the common contract used by all supported engines for engine detection, typed data loading, required-file resolution, and engine-appropriate writing. Its *name* property identifies the adapter (i.e., ReaxFF, AMS, or LAMMPS), while *detect* is defined as an abstract method, meaning subclasses such as ReaxFF, LAMMPS, or AMS adapters must provide their own detection logic. In ReaxKit, *detect* should return a confidence score from 0 to 1, allowing the runtime to determine whether a given path likely contains files generated by a particular engine.

The *load* method behaves as a factory-like dispatcher: when a caller requests a specific type of data (such as trajectory data), this method calls the appropriate engine-specific loader method (here, *load_trajectory*). In other words, It centralizes type-to-loader routing while delegating the actual loading behavior to concrete adapter methods. Once the data is loaded the Engine-specific loader method, it will be mapped to canonical domain model classes, whose full list is shown in Figure S2. This mapping to canonical data classes is what makes engine support scalable: different engines can produce different raw outputs, but ReaxKit converts them into the same internal data structures, allowing downstream analysis, visualization, and workflows to remain engine-independent. The *write* method follows the same pattern in the opposite direction, routing canonical domain objects, such as *TrajectoryData* or *ControlParametersData*, to engine-specific writers, for writing trajectories or control parameters into *xmolout* and *control* files, respectively.

Under *src/reaxkit/engine/*, there are dedicated subpackages for *reaxff/*, *lammps/*, and *ams/* engines, with shared functions placed in *common/*. Then, each engine folder contains the following main components:

- *I/O* handlers for parsing native simulation files (i.e., *xmolout_handler.py*),
- *generators* for producing engine-compatible inputs (i.e., *trainset_generator.py*),
- adapter modules to connect these file-level operations to canonical ReaxKit data structures (i.e., mapping *xmolout* data to the domain model *TrajectoryData*),

- and a *data/* folder providing engine-specific reference and documenting how native engine outputs are organized and mapped into ReaxKit's canonical classes.

## 2.3. Analysis modules

ReaxKit supports multiple analyzer modules which are categorized by their functionality into such active sites, connectivity, control, electrostatics, force field, kinematics, molecular analysis, force field optimization params, timeseries, and trajectory folders. Because the engine layer normalizes ReaxFF, LAMMPS, AMS, or other supported outputs into canonical domain models, the analysis modules can focus on scientific computation rather than file-format handling.

All analyzer follows a consistent Request–Task–Result pattern as shown in **Figure 2**: a *request* dataclass stores user-configurable options, a *task* class performs the computation, and a *result* dataclass stores the computed output together with the request used to generate it. In this figure, *required_data*, *run*, and *recommended_presentation* are colored the same since they are all implemented under the *task*. For a template analyzer, *TemplateRequest* inherits from *BaseRequest* and contains configurable fields such as selected atom or entity IDs, selected labels, dimensions, analysis mode, frame indices, stride, reference options, thresholds, and other settings. These fields also include metadata used for labels, help text, choices, units, and UI rendering, consistent with the use of dataclass field metadata as an extension mechanism in Python. The task class, *TemplateTask*, inherits from *AnalysisTask*, is registered with a task name and label, declares its required input data type as *required_data*, and implements a *run(data, request)* method that returns a typed *TemplateResult*. The result object inherits from *BaseResult* and stores both the output table and the original request, preserving analysis provenance and enabling consistent downstream presentation.

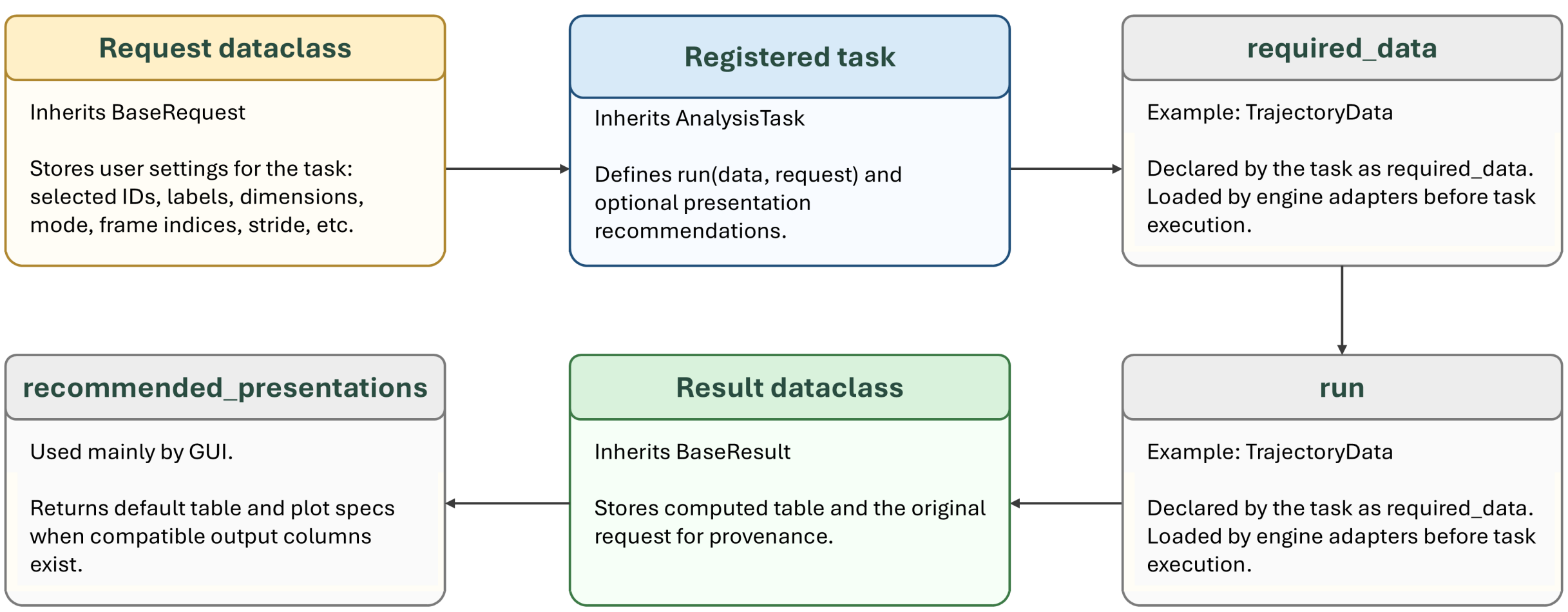

**Figure 2: Request–Task–Result pattern used by ReaxKit analysis modules.** Each analyzer is implemented as a typed request object, a registered analysis task, and a typed result object. The task declares the canonical data required for execution, runs the analysis using user-provided request settings, and returns structured results that can be rendered as tables, plots, files, or graphical-interface views.

## 2.4. Workflow automation

ReaxKit workflow modules provide the orchestration layer that connects user-facing commands to registered commands (i.e., analysis tasks or generators). In workflows, the CLI entrypoint is responsible for startup, path bootstrapping, and initial command dispatch, while workflow modules define the command-specific behavior needed to run commands. Figure S3 illustrates this structure through two central functions: *build_parser()* and *run_main()*. The *build_parser()* function configures the command-specific argparse parser, resolves canonical command names from CLI aliases, defines user-facing options such as engine selection, run directory, atom filters, frame selection, plotting, saving, and export settings, and adds common storage-related arguments. After parsing, workflow execution proceeds through *run_main()*, which resolves the command name, retrieves the appropriate task class from the task registry, builds a typed request object from the parsed arguments, executes the task through the *AnalysisExecutor* (will be discussed shortly), and sends the result to the presentation/export layer.

Two other important components within workflows execution are *command registry* and *task registry*, making ReaxKit workflows extensible and maintainable. The *command registry* normalizes user-facing command names and aliases into canonical command identifiers, allowing the CLI to support both stable command names and backward-compatible or user-friendly aliases without duplicating workflow logic (for example, users can use both *msd* or *mean-square-* *displacement* as the command name when writing a CLI command). Once the command is resolved, the *task registry* maps the canonical command name to the corresponding analysis task class, decoupling command parsing from task implementation. As a result, ReaxKit can support a growing set of analysis workflows while preserving a consistent execution pattern: parse the command, resolve the task, build the request, execute the task through the analysis executor, and dispatch the result for presentation or export.

### 2.4.1. Request-to-Result Pipeline through the Analysis_executor

As mentioned earlier, once the user intent is translated into a task definition and a structured request object, the execution will be delegated to *AnalysisExecutor*. This simplifies task execution by centralizing the execution logic in *AnalysisExecutor*, so each workflow only needs to specify the requested task while the executor manages how it is run, leading to a separation of concerns. **Figure 3** shows how *AnalysisExecutor* routes users requests and obtains the results.

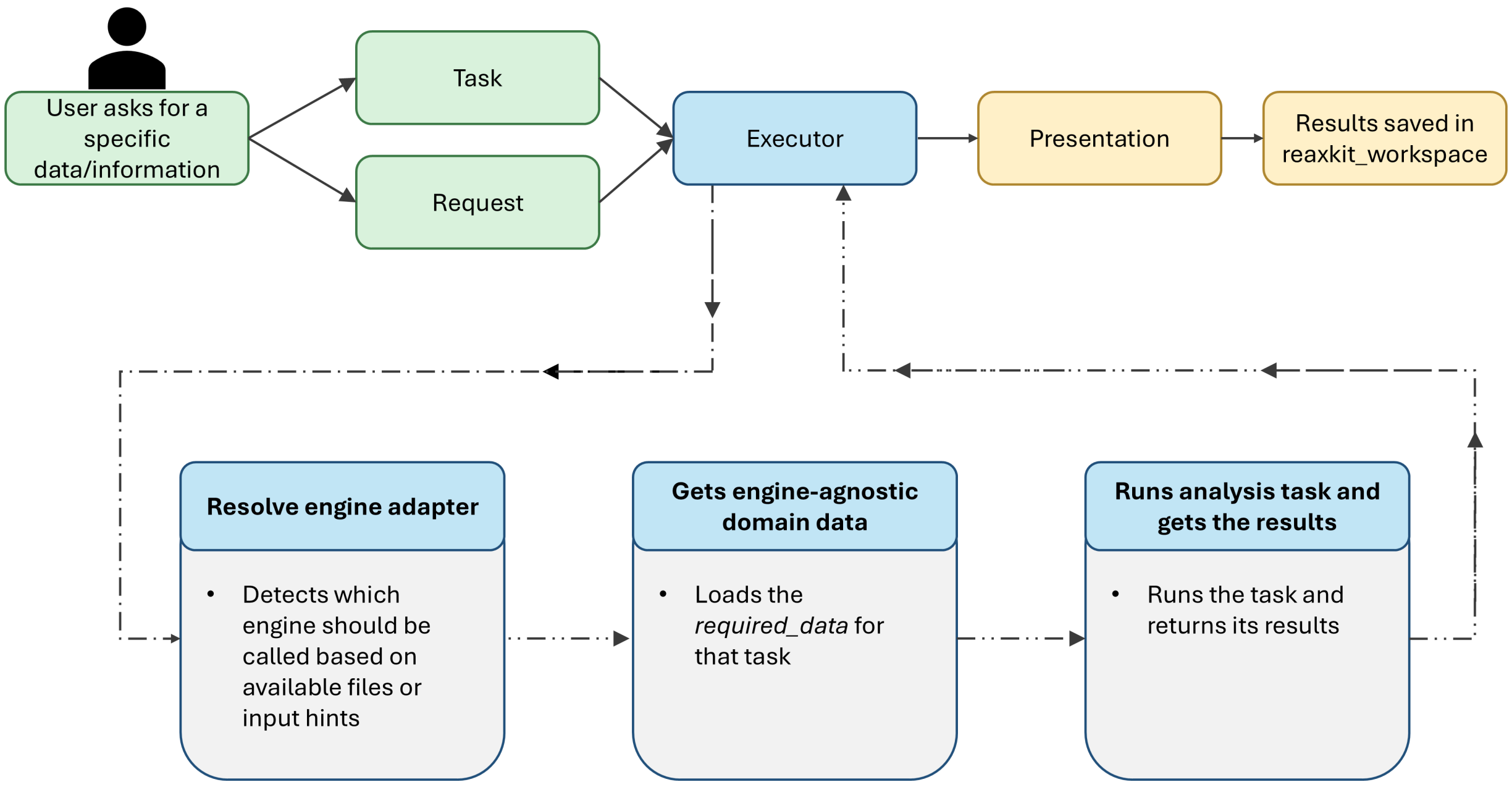

**Figure 3: Engine-independent request-to-result workflow in ReaxKit.** User requests from the CLI (or graphical interface) are converted into task and request objects, routed through an executor, and applied to normalized data structures. The resulting analysis objects are passed to the presentation layer and can be saved in the ReaxKit workspace, providing a consistent workflow across different engines.

*AnalysisExecutor* mainly does 3 tasks:

1. Resolve engine adapter from input hints and snapshot required raw,

2. Asking for the task's *required_data*, and loading it as domain data classes,
3. Runs the task and gets the results back.

Once the results are obtained, presentation layer helps with saving, plotting, or exprting them. As an example, consider the following case where user wants to do his first reaxkit call in a folder with ReaxFF-generated files:

1. If the user already knows the appropriate CLI command and its flags, they can execute it directly in the terminal. Otherwise, ReaxKit can assist in identifying the appropriate command through a query such as *reaxkit help "get msd"*. Here, the text enclosed in quotation marks represents the user's query and expresses the intended task. ReaxKit's search and help modules interpret this intent and return relevant CLI commands, together with information about the required data, input files, and analyzers. For example, for the query *"get msd"*, ReaxKit would identify *get_msd* as the relevant command and provide the information needed to use it. To view the available flags for *get_msd*, the user can then execute *reaxkit get_msd -h*. This displays a brief description of the command, several useful and commonly used example invocations, and a table summarizing all available flags. For each flag, the help output indicates whether it is mandatory or optional, explains its purpose, and provides its default value where applicable.
2. User writes the CLI command *reaxkit get_msd --atom-ids 1 2 3 --max-lag 500 --delta-t-ps 0.25 --plot single*, where he asks for calculating MSD (i.e., the task) of atoms 1, 2, and 3 with specific settings (forming the task request), and then plotting the results on a single figure (i.e., all plots sharing the same axes).
3. Once the intent of user is clarified, *analysis_executor* is called to ask for the MSD's *required_data* (which is *TrajectoryData*).
4. Then, atomic trajectory data is loaded by *load_trajectory* method within the ReaxFF adapter, reading a *xmolout* file.
5. Loaded data is then mapped to *TrajectoryData*, and is used by MSD task to calculate the mean square displacement with the given settings.
6. Finally, the results are sent to the presentation layer for plotting the results in a single figure.

In addition to these core responsibilities, *AnalysisExecutor* also provides several supporting functions that make ReaxKit executions more reproducible, traceable, and efficient. It normalizes runtime and storage arguments, derives task and data metadata, and initializes runtime side effects such as terminal logs, file logging, timing hooks, cache settings, and log verbosity. When a run-scoped folder layout is active, it also registers dataset and run metadata so that parsed inputs, task settings, and generated outputs can be associated with a specific execution. Before loading or recomputing data, *AnalysisExecutor* checks whether the required parsed domain data or previous analysis results are already available in cache, allowing repeated tasks to use a faster execution path when possible. After the task is completed, it enriches the returned result with useful metadata, such as timing or time-axis information, and persists the relevant cache and log records. In this way, AnalysisExecutor does not only run the requested analysis; it also manages the execution environment around the task, making the workflow more organized, repeatable, and easier to debug.

## 2.5. Utility functions

The *utils/* layer contains cross-package helper functionality that supports ReaxKit without belonging to a specific engine, analyzer, workflow, or presentation module. It includes reusable numerical and other utilities (like equation-of-state helper functions), and is intended to be imported by generator, analysis, workflow, or presentation modules when generic helper logic is needed. Because this layer is designed for low-coupling support code, its functions follow a pure-function style: they receive explicit inputs, return computed outputs, and avoid hidden dependence on global state or workflow-specific side effects. This makes utility functions easier to test, reuse, and combine across different parts of ReaxKit.

## 2.6. Presentation layer

The presentation layer is responsible for converting analysis results into user-facing outputs after the scientific computation has already been completed. In ReaxKit, this layer contains report payload builders, presentation specifications, conversion and dispatch helpers, persistence/export utilities, unit-formatting helpers, and integrations for plot and movie rendering. Its directory structure separates specialized output logic into subpackages *plot/* for plot registry and renderer integration supporting different types of plots (i.e., beeswarm, boxplot, dual y-axis, etc.), *movie/* for video-generation helpers, and a specific report generator for *active_sites/* for active-site analysis (other analyzers can have their own reporting as well). Meanwhile, modules such as *dispatcher.py, specs.py, reporting.py, persist.py, convert.py, units.py, and export_utils.py* handle generic presentation and export behavior.

In the overall ReaxKit workflow, the analysis layer returns typed result objects or serialized result payloads, and the presentation layer determines how those outputs should be displayed, saved, or converted into artifacts such as tables, figures, reports, movies, or persisted output files. This design allows analysis tasks to define or recommend presentation specifications without directly depending on plotting or export code, while workflows and core runtime components can pass result payloads into common presentation utilities for consistent rendering and storage. As a result, ReaxKit separates "what is computed" from "how it is shown," making it easier to add new renderers, report builders, or export formats without changing the underlying analysis or engine-adapter code.

### 2.6.1. ReaxKit Workspace for Organized and Reproducible Outputs

To keep generated files and intermediate artifacts organized, ReaxKit uses a dedicated *reaxkit_workspace* directory as the default storage location for generated input files, simulation data, analysis outputs, caches, logs, and reports, ass shown in **Figure 4**.

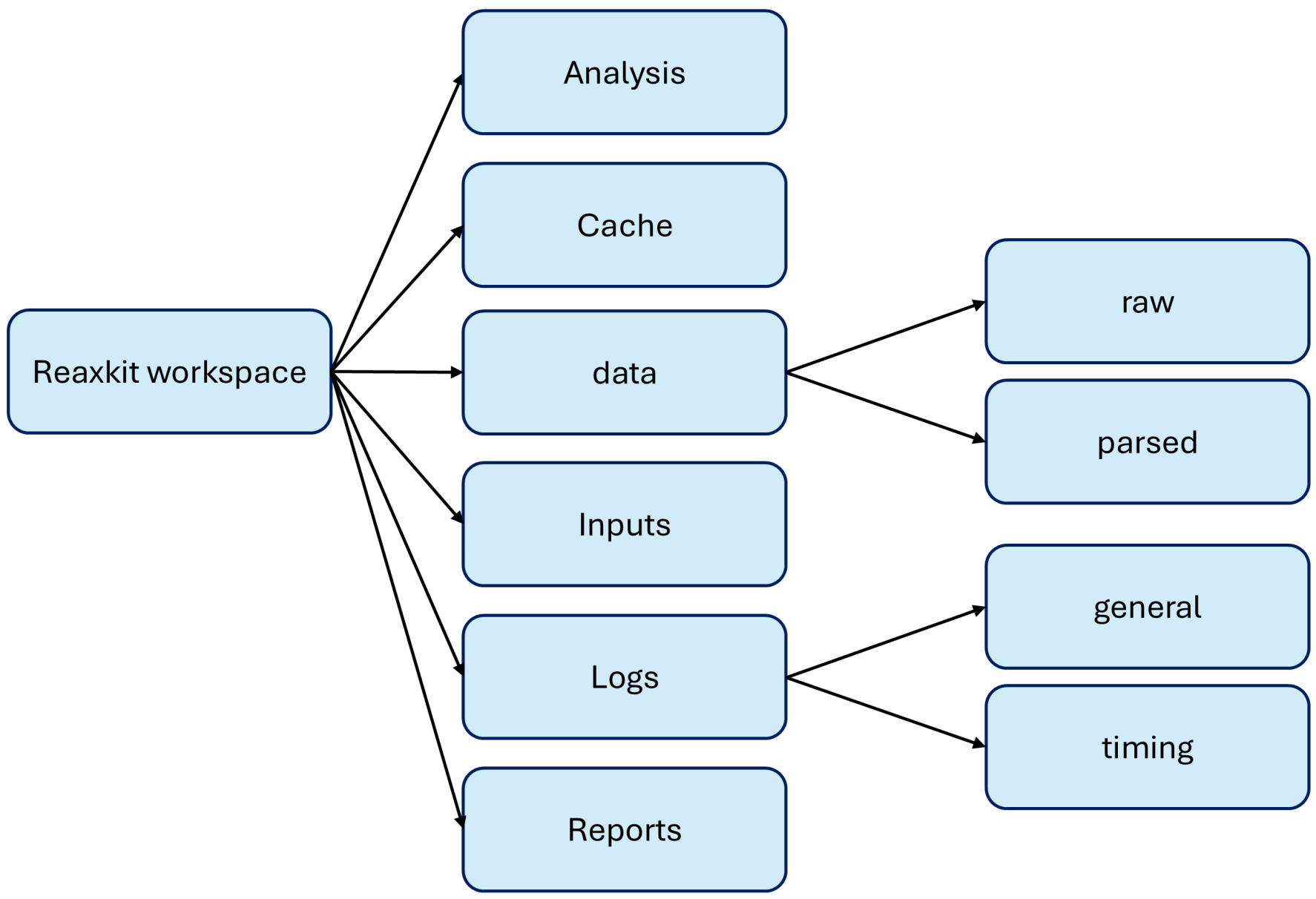

**Figure 4: Overall folder structure of the *reaxkit_workspace*.** ReaxKit organizes generated artifacts into dedicated folders for analysis outputs, cached objects, raw and parsed data, generated inputs, logs, and reports. The workspace separates original simulation files from normalized data and analysis results while preserving traceability through identifiers such as *run_id, parsed_id, and analysis_id*.

ReaxKit organizes files according to their role in the simulation and analysis workflow. ReaxFF input files generated by ReaxKit, such as a *control* file with user-defined simulation settings, are stored under *inputs/<run_id>/*. Raw ReaxFF simulation output files, such as trajectory, charge, or summary files produced by a completed ReaxFF run, are stored under data/raw/<run_id>/, where each run-specific folder contains the original files required by the requested task. After these raw files are loaded and normalized into ReaxKit domain models, the corresponding processed artifacts are written under *data/parsed/<parsed_id>/* as serialized data and metadata files. Analysis results, including computed tables and figures, are stored under *analysis/<task>/<analysis_id>/*, and reports with WORD or PDF format are saved under *reports/<analysis_id>/*. Additional folders are reserved for supporting execution artifacts, such as reusable caches, logs, and timing records. This structure allows users to clearly distinguish generated simulation inputs, raw ReaxFF outputs, normalized data, computed analysis results, and auxiliary execution records.

This workspace is designed to preserve traceability between raw files, parsed datasets, analysis results, and generated outputs. ReaxKit uses several identifiers for this purpose. A *run_id* identifies a specific simulation or task run and is generated from the current timestamp plus a short random token in the format run_YYYYMMDD_HHMMSS_<token>. A *parsed_id* identifies normalized data produced from a raw snapshot; it is generated as a short SHA-256 hash from the raw snapshot fingerprint and handler version, meaning that identical raw inputs processed with the same handler version can resolve to the same parsed identity even if their run identifiers differ. An *analysis_id* identifies a specific analysis execution and is generated from the parsed-data identity, task class, task version, and request fingerprint; when no parsed identifier is available, ReaxKit falls back to hashing the task, data, and request objects. This makes the output directory structure both human-navigable and machine-traceable where each analysis result can be traced back to the raw files, the normalized data object, and the request parameters used to produce it. It also improves practical usability: after a task is executed, ReaxKit reports where the relevant files were saved, making it easier to locate generated inputs, parsed data, plots, tables, logs, or reports. The cache layer further supports repeated or high-throughput workflows by storing reusable handler and analysis outputs, while the log and timing folders provide execution traces that can be used for debugging and performance inspection.

Below shows some example file placements:

- **inputs**: *inputs/<run_id>/control* is a generated control file
- **data/raw**: *data/raw/<run_id>/xmolout* is a snapshot of the original ReaxFF output file (i.e., *xmolout*) copied into the workspace for a specific run.
- **data/parsed**: *data/parsed/<parsed_id>/trajectorydata.h5* and *data/parsed/<parsed_id>/meta.json* are the ReaxKit's normalized and structured representations of the raw data, stored in HDF5 (*Hdf5*, n.d.) and JSON formats for efficient access and metadata storage.
- **data/run_index**: *data/run_index/<run_id>.json* is a JSON file that serves as an index for all the runs you have executed, containing metadata about each run such as the files used, parameters, and timestamps.
- **analysis**: *analysis/msd/<analysis_id>/result.csv* and *analysis/msd/<analysis_id>/settings.json* are the outputs of an MSD analysis task, where *result.csv* contains the computed mean squared displacement values and *settings.json* contains the parameters and configuration used for that analysis.

- **cache/handlers**: *cache/handlers/<handler_id>/cache.h5* is a cache file that stores intermediate data and results related to a specific handler (i.e., *XmoloutHandler*), allowing for faster access and reuse of previously computed information when the same handler is invoked again with the same parameters.

- **cache/analysis**: *cache/analysis/<analysis_id>/cache.h5* is a cache file that stores intermediate data and results related to a specific analysis task (i.e., *MSDTask*), enabling efficient retrieval of previously computed results when the same analysis is performed again with the same parameters.

- **cache/index**: *cache/index/handlers.json* and *cache/index/analysis.json* are JSON files that serve as indexes for the handler and analysis caches, respectively, keeping track of which cache files correspond to which handlers and analyses, along with metadata such as timestamps, parameters, and relationships between different cached artifacts.

- **logs**: *logs/general/reaxkit_general.log* and *logs/timing/human_readable_timing.log* are log files that capture general logs and timing information for your ReaxKit runs, providing insights into the execution flow, performance, and any issues that may arise during the processing of your data and analyses.

- **reports**: *reports/<analysis_id>/active_site_events_report.pdf* is a generated report that summarizes the findings of a specific analysis task (i.e., active site events), presenting the results in a structured and visually appealing format for easy interpretation and sharing.

## 2.7. Graphical user interface using Dash

The WebUI layer provides a browser-based interface for interacting with ReaxKit without requiring users to directly write command-line arguments or Python scripts. It is implemented as a Dash-based (*Dash*, n.d.) web application layer, with separate components for application wiring, callbacks, runtime paths, UI composition, backend serialization, and web-specific presentation behavior. The webui package is organized into *backend/*, which handles API schemas, serializers, and backend registries; *ui/*, which contains page layouts and callback-facing UI components; and *presentation/*, which manages web presentation registry and performance-related settings. This layer does not implement engine parsing or analysis algorithms; instead, it issues command requests, receives analyzed or persisted payloads from the ReaxKit runtime, and renders interactive views and export actions for the user.

The ReaxKit WebUI is intended to run locally rather than as an externally deployed web service. Users start it with the CLI command *reaxkit gui*, after which the Dash application is served on localhost and accessed through a web browser. This design choice keeps simulation data and analysis outputs on the user's own machine or within the computational environment where the simulations are being analyzed. It is especially useful for interactive desktop sessions on high-performance computing clusters, where users often want the WebUI to use the same allocated resources and file system context as their active job rather than transferring data to, or requesting resources from, an external server. In this role, the WebUI provides an accessible graphical layer over the same ReaxKit workflow, analysis, presentation, and export infrastructure while preserving local execution, data locality, and compatibility with workstation or HPC-based research workflows.

**Figure 5** illustrates the ReaxKit GUI, which provides a pipeline-based environment for loading simulation datasets, configuring analyses, and visualizing results. Users interact primarily with the *Pipeline Browser* on the left, where a dataset is loaded and ReaxKit attempts to automatically identify the simulation engine and associated input and output files. If automatic detection fails because of nonstandard file names or directory organization, the user can manually specify the engine and relevant file paths through the *Engine* component of the pipeline. In the example shown in **Figure 5**, the ReaxFF engine has been detected and 1551 simulation frames have been loaded, as reported in the bottom *status bar*.

Analysis modules can then be added to the pipeline and configured for the desired task. In this example, a radial distribution function (RDF) analysis has been executed. The resulting data can subsequently be passed to presentation components and displayed in the right-side *Visualization Canvas* as plots, tables, or other supported representations; here, the calculated RDF is shown as a two-dimensional plot of $g(r)$ versus $r$. The lower-left *settings panel* dynamically updates according to the currently selected pipeline component, allowing users to modify dataset options, analyzer parameters, or presentation settings such as the plotted variables. This pipeline-based organization therefore provides a visual representation of the sequence from simulation data, through analysis, to presentation of the resulting data.

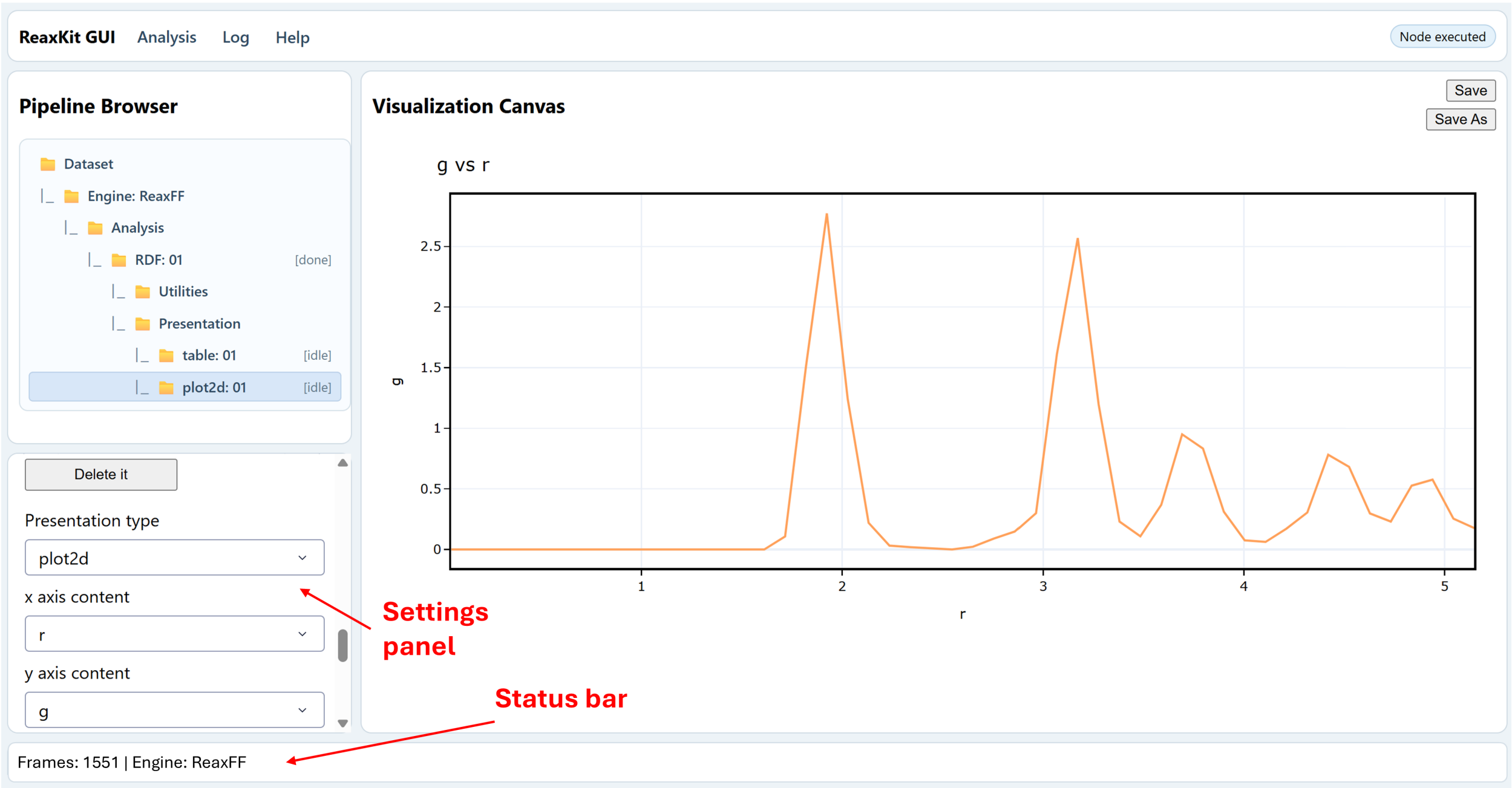

**Figure 5: ReaxKit graphical user interface.** The WebUI provides a local, browser-based interface for loading simulation datasets, configuring analysis pipelines, and visualizing results. The left panel contains the *pipeline browser* and component-specific *settings panel*, the central panel provides the *visualization canvas* for tables and plots, and the bottom *status bar* reports the current dataset and engine-loading state.

## 2.8. Supporting Components and Package Organization

Beyond the modules discussed above, ReaxKit also includes several supporting components that make the package easier to use, extend, and maintain. Collectively, all these modules are organized ReaxKit as shown in Figure S4. Under *src/reaxkit/*, the *cli, core, data,* and *domain* packages form the foundational layers of ReaxKit. The *cli* layer provides the user-facing command-line interface and serves as a higher-level entry point for running ReaxKit workflows: it receives user commands, parses command-line options, prepares paths and runtime settings, and then dispatches the request to the appropriate workflow.The *core* layer provides the execution infrastructure, including command and task registries, alias resolution, caching and storage management, and runtime coordination. The *data* layer contains shared static information used across the package, such as constants, units, variable aliases, and metadata. The *domain* layer defines ReaxKit's data contracts, including typed request objects, result objects, and canonical data models.

The *help* package provides ReaxKit's help and discovery system. It loads developed YAML metadata describing commands, scientific intents, file meanings, aliases, workflows, generators, analyzers, and data mappings. This allows users to search within ReaxKit to find the relevant generators, analysis modules, files, or workflows related to their intent. For example, if users want to know how to get the MSD, they can simply write the CLI command *reaxkit help "msd"*. The package also includes introspection utilities that scan Python modules and extract short descriptions from docstrings, allowing ReaxKit to partially describe its own capabilities

# *3.* Representative ReaxKit Capabilities

After describing the overall architecture and package organization, this section highlights representative ReaxKit capabilities across different stages of a ReaxFF workflow:

- Sections 3.1, Training Set Generation, and 3.2, Force-Field Merging, demonstrate ReaxKit's capabilities for generating and preparing input files.
- Section 3.3, Force-Field Optimization Diagnostics, focuses on processing and interpreting output files produced during force-field optimization rather than molecular dynamics simulations, whereas 3.4, Active Site Analysis, illustrates analysis of MD simulation outputs.
- Section 3.5, Regeneration of Simulation Files Based on Derived Atomic Properties, combines analysis and file generation by using derived simulation properties to construct new trajectory files.
- Finally, 3.6, Study Design, demonstrates ReaxKit's workflow-automation capabilities for organizing and executing systematic simulation studies.

## 3.1. Training set generation

Training sets are a central component of force-field optimization because they define the reference physical and chemical behavior that the fitted force field is expected to reproduce. In ReaxFF development, a training set typically contains quantum-mechanical or experimental reference data for representative structures, reactions, and deformation pathways, including energies, bond dissociation curves, equations of state, atomic charges, reaction barriers, and others.

### 3.1.1. Training set generation using mechanical data

One important training set is energy–volume data, or equation-of-state data, which is generated using material's mechanical properties, such as elastic constants and stiffness tensors. This data helps the optimized force field capture both structural stability and mechanical deformation behavior. The elastic properties of solid-state materials are fundamental parameters that bridge microscopic interatomic interactions with macroscopic mechanical behavior. Thermodynamically, elastic constants exhibit strong correlations with key thermal properties, including specific heat capacity and thermal expansion coefficients. Furthermore, empirical and theoretical evidence establishes a direct link between a solid's elastic moduli and its melting point, highlighting their importance in high-temperature material stability. Mechanically, the complete tensor of elastic constants dictates the structural response to external forces, directly determining the macroscopic bulk, shear, and Young's moduli, as well as Poisson's ratio, hardness, and overall mechanical strength.

Although density functional theory (DFT) calculations offer a robust route to computing these mechanical properties from first principles, the computational cost escalates dramatically when searching across vast chemical spaces or dealing with low-symmetry crystal structures. To circumvent the prohibitive expense of running exhaustive first-principles calculations from scratch, high-throughput reference databases such as the Materials Project can be leveraged as an efficient starting point. The Materials Project serves as a comprehensive, open-access repository providing pre-computed equilibrium crystal structures and elastic tensors. However, while these static database values are invaluable, they do not inherently provide the comprehensive off-equilibrium configurations, strained geometries, and corresponding energy-strain trajectories required to train robust reactive force fields. To bridge this data gap, we developed a new automated scripting workflow designed to systematically parse reference data from the Materials Project and autonomously generate extensive energy-strain and geometry datasets. By automating the application of diverse distortion matrices – specifically tailored for orthorhombic crystal structures – this workflow eliminates the manual bottleneck of setting up independent deformed configurations. This automated pipeline ensures a rapid, high-throughput generation of self-consistent training sets, significantly accelerating the development and deployment of accurate force fields for mechanical property predictions.

## Theoretical framework and energy-strain formalism

The continuum mechanics framework relates a material's thermodynamic energy directly to microscopic lattice deformations. In this context, strain describes the relative displacement of atomic positions that alters the material's volume, shape, or both. These deformations are broadly categorized into isotropic strain (equal deformation across all three dimensions), uniaxial strain (deformation restricted to a single axis), and volume-conserving configurations such as pure shear strain. To evaluate the elastic constants, the internal energy $E(\delta)$ of a crystal subjected to a small strain tensor is expressed via a Taylor expansion about its equilibrium state as shown by Eq. (1):

$$\boldsymbol{E(\delta) = E_0 + V_0\left(\sum_{i=1}^{6} \tau_i \epsilon_i + \frac{1}{2}\sum_{i=1}^{6}\sum_{j=1}^{6} c_{ij}\epsilon_i\epsilon_j\right) + \mathcal{O}(\delta^3)} \quad \textbf{(1)}$$

where $E_0$ and $V_0$ represent the energy and volume of the unstrained equilibrium lattice, respectively; $\tau_i$ represents the components of the initial stress tensor; $\epsilon_i$ and $\epsilon_j$ are the strain components expressed in Voigt notation; and $c_{ij}$ denotes the second-order elastic constants. Because the initial reference structure is fully relaxed to its equilibrium state at zero external pressure, the first-order residual stress terms ($\tau_i$) vanish. Under the assumption of sufficiently small applied strains, third-order and higher-degree terms ($\mathcal{O}(\delta^3)$) become negligible. Consequently, the second-order elastic constants are directly proportional to the quadratic coefficients obtained from a polynomial fit of the calculated total energy as a function of the applied distortion strain. This mathematical relationship forms the basis of our automated workflow, which samples discrete energy-strain curves to extract the full elastic tensor.

## Application to orthorhombic symmetries and distortions

The number of independent elastic constants required to fully characterize a crystalline material is strictly governed by its spatial crystal symmetry. Consequently, automated configuration generator within this workflow is tailored for application to orthorhombic crystal structures. Defined by three mutually orthogonal but unequal lattice parameters, the elastic tensor of an orthorhombic system reduces to nine independent components: $c_{11}$, $c_{22}$, $c_{33}$, $c_{44}$, $c_{55}$, $c_{66}$, $c_{12}$, $c_{13}$, and $c_{23}$. To systematically isolate these individual coefficients, a symmetric distortion matrix $D$ – parameterized according to standard volume-conserving or directional strains (Ravindran et al., 1998) – is applied to the equilibrium lattice vectors, mapping the macroscopic deformation directly onto the underlying atomic coordinates.

When the orthorhombic crystal undergoes controlled elastic deformation via a specific directional strain ($\delta$), the energy-strain relationship simplifies to localized quadratic expressions. For instance, applying a uniaxial strain along the principal $x$-axis yields an energy variation defined by Eq. (2):

$$\boldsymbol{E(\delta) = E_0 + \frac{c_{11}}{2} V_0 \delta^2} \tag{2}$$

To isolate the off-diagonal elastic components ($c_{12}$, $c_{13}$, and $c_{23}$), a distortion matrix combining multi-axial coupled strains is implemented. Under this asymmetric configuration, the energy-strain relationship expands to incorporate cross-coupling terms, such as Eq. (3):

$$\boldsymbol{E(\delta) = E_0 + \frac{V_0}{2}(c_{11} + c_{22} - 2c_{12})\delta^2} \tag{3}$$

Similarly, the pure shear constants ($c_{44}$, $c_{55}$, and $c_{66}$) are independently resolved through volume-conserving shape deformations of the simulation cell via the expression in Eq. (4):

$$\boldsymbol{E(\delta) = E_0 + 2V_0 c_{44} \delta^2} \tag{4}$$

By evaluating the total energy across a series of incremental steps for each independent distortion matrix, the full orthorhombic elastic tensor can be robustly resolved. Knowing the complete elastic tensor allows for a self-consistent workflow where the strain energy can be predicted for any arbitrary deformation matrix.

Finally, isotropic volume deformations are systematically applied to generate an independent energy-volume dataset. These data points are fitted to the third-order Birch-Murnaghan equation of state (Birch, 1988) to precisely extract the material's bulk modulus ($B_0$). Because the Materials Project database does not provide the first derivative of the bulk modulus with respect to pressure ($B_0'$), a default empirical value of 1.5 is adopted within the script to prevent an unphysical, completely symmetric E-V curve. While this approximation introduces a localized mathematical discrepancy regarding the curvature asymmetry, the resulting error remains negligible under the small-strain regime utilized here. To ensure the validity of this approximation while capturing sufficient curvature data, the workflow restricts the isotropic deformations to a narrow window of volume change. The extracted $B_0$ value ultimately serves as a self-consistent cross-check against the bulk modulus derived analytically from the individual elastic constants, confirming the physical reliability of the generated training dataset.

**Workflow implementation and training set generation**

We have implemented the necessary modules in ReaxKit to generate such data using Material's project website for any number of materials. The practical pipeline for generating the training dataset begins by retrieving reference crystal structures and their corresponding elastic tensors. These baseline structures are converted into standard .xyz atomic coordinate format to serve as the primary inputs for the automated generation script. The script parses these initial geometries alongside the reference elastic data to construct a master training data file, designated as trainset.in. By systematically applying the parameterized distortion matrices described previously, the automated pipeline generates a comprehensive ensemble of strained configurations. The definitive outputs of this workflow comprise the initialized trainset.in, structured energy-versus-strain datasets optimized for plotting, and the complete set of deformed atomic geometries exported in both .geo and .xyz formats for direct integration into force field training loops.

A simple way to use the training-set generator is through the command-line interface. For example, the command *reaxkit gen_elastic_trainset --input-mode batch --elements Ba,B,O --api-key YOUR_KEY* generates energy–volume training data for Materials Project compounds that contain Ba, B, and O and have available mechanical-property data. ReaxKit retrieves the required structural and elastic information through the Materials Project API (Jain et al., 2013), uses these data to construct equation-of-state training entries, and writes the resulting training-set files in a format suitable for force-field optimization.

Once the user executes the above-mentioned CLI command, ReaxKit generates per-material structures and volume–energy data, along with the corresponding training-set and *geo* files that can be directly used for force-field optimization. In addition, ReaxKit writes a *materials_status.csv* file, an example of which is shown in **Table 1**. This file summarizes the outcome for each candidate material retrieved from Materials Project, including its chemical formula, crystal system, Materials Project ID, processing status, and any warning messages. In this example, most candidate structures are marked as *skip* (since there was no mechanical data available for these materials), while one structure, $Ba(BO_2)_2$ with material ID mp-5730, is marked as *success*. The warning column provides additional diagnostic information when needed; for example, it notes that the successful structure has a non-orthogonal bulk and elastic cell, meaning that the generated elastic energy targets should be interpreted with this lattice assumption in mind. This status table allows users to quickly identify which materials were successfully converted into training-set entries, which were skipped, and why additional inspection may be required.

**Table 1: Example *materials_status.csv* output from the ReaxKit elastic training-set generator.** The table reports candidate Ba–B–O materials retrieved from Materials Project, including their chemical formula, crystal system, Materials Project ID, processing status, and warning messages. It allows users to identify which structures were successfully converted into training-set entries, which were skipped, and whether any generated entries require additional inspection.

| Chemical formula | Crystal system | Material ID | Status | Warning |
|---|---|---|---|---|
| $Ba(BO_2)_2$ | Monoclinic | mp-1182856 | skip | |
| $BaB_2O_9$ | Monoclinic | mp-1196854 | skip | |
| $Ba(BO_4)_2$ | Monoclinic | mp-1201084 | skip | |
| $Ba(BO_2)_5$ | Triclinic | mp-1204604 | skip | |
| $BaBO_3$ | Cubic | mp-1214407 | skip | |
| $Ba(BO_2)_2$ | Trigonal | mp-5730 | success | Bulk cell is non-orthogonal. |
| $BaB_4O_7$ | Monoclinic | mp-27692 | skip | |
| $BaB_8O_{13}$ | Tetragonal | mp-27794 | skip | |
| $Ba(BO_2)_2$ | Trigonal | mp-540659 | skip | |
| $BaB_4O_7$ | Orthorhombic | mp-556974 | skip | |
| $Ba(BO_2)_5$ | Triclinic | mp-728866 | skip | |
| $Ba_2B_2O_5$ | Monoclinic | mp-768345 | skip | |
| $Ba_2B_2O_5$ | Monoclinic | mp-771158 | skip | |
| $Ba_2B_2O_5$ | Monoclinic | mp-771174 | skip | |
| $Ba_2B_{10}O_{17}$ | Triclinic | mp-1019522 | skip | |
| $Ba_2B_5O_{11}$ | Monoclinic | mp-1198603 | skip | |
| $Ba_2B_2O_5$ | Triclinic | mp-753413 | skip | |
| $Ba_3(BO_3)_2$ | Orthorhombic | mp-779608 | skip | |
| $Ba_3(BO_3)_2$ | Trigonal | mp-755417 | skip | |
| $Ba_5B_4O_{11}$ | Orthorhombic | mp-3974 | skip | |

### 3.1.2. Automated Isomer-Based ReaxFF Training-Data Generation

A three-stage workflow was implemented in ReaxKit to automate the preparation of ReaxFF training data from molecular dynamics trajectories. Previously, this process required several manual steps: inspecting trajectory outputs, identifying candidate molecular isomers, preparing and submitting individual DFT jobs for each isomer, and finally extracting optimized geometries and relative energies into ReaxFF-compatible training files. The implemented workflow formalizes these operations into reproducible ReaxKit commands while keeping each stage independently executable.

The overall workflow consists of: 1. isomer representative detection from ReaxFF MD outputs, 2. quantum chemical simulation job preparation and optional submission, and 3. generation of ReaxFF training-set data from completed quantum outputs. An overview of the complete three-stage workflow and the data exchanged between the individual stages is shown in **Figure 12**.

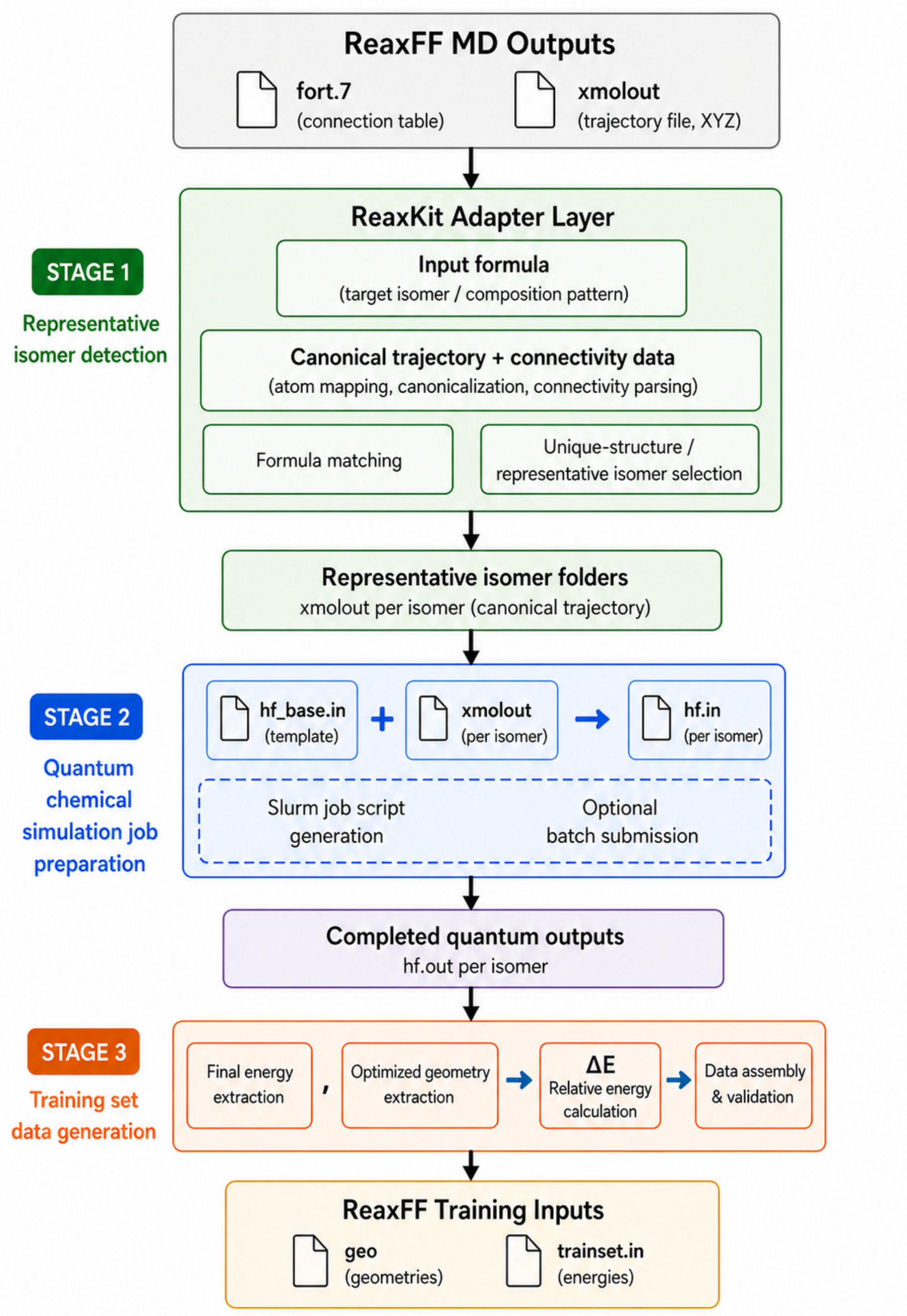

**Figure 10. Overview of the three-stage ReaxKit workflow for representative isomer identification and automated ReaxFF training-data generation.** Stage 1 identifies representative isomers from ReaxFF MD trajectories, Stage 2 prepares and optionally submits quantum chemical simulation jobs, and Stage 3 extracts optimized geometries and relative energies to generate ReaxFF-compatible training inputs.

## Stage 1: Representative Isomer Detection

The first stage identifies representative isomer candidates from ReaxFF MD trajectory (i.e., *xmolout*) and connectivity (i.e., *fort.7*) data. Representative isomers are selected using a coarse same-formula criterion. The purpose is not to exhaustively enumerate all graph-distinct molecular isomers, but to obtain a compact set of chemically relevant structures sampled during the simulation. This is important for ReaxFF training because downstream quantum calculations are computationally expensive, and using every similar trajectory structure would add cost without necessarily adding proportional training value. The workflow therefore supports limiting the maximum number of representatives, allowing users to control the number of structures passed to quantum calculations. The output of this stage is a set of representative structures, optionally written as one folder per selected isomer. These folders provide the handoff point to the next stage.

**Stage 2: Quantum chemical simulation job preparation and submission**

The second stage automates the preparation of quantum chemical simulation jobs for the selected isomer representatives. The current implementation supports Jaguar (Schrödinger Inc, USA), which is used for the quantum-chemistry calculations. For each selected isomer, ReaxKit prepares the corresponding Jaguar input by combining a user-provided *hf_base.in* template with the representative molecular geometry. The generated input files are organized into individual isomer directories, providing a systematic interface between the representative structures identified in stage 1 and the subsequent quantum calculations.

The workflow also generates Slurm-compatible job scripts for each isomer. Cluster-specific values such as partition, wall time, memory, number of CPUs, module loading, and executable command are provided through configuration rather than hardcoded into the workflow. This is necessary because HPC environments differ across institutions and systems. The current implementation supports Slurm submission through sbatch, while the code structure leaves room for additional scheduler backends in the future. The job-generation stage can optionally submit jobs automatically. It also includes safeguards for practical high-throughput use: existing queued jobs can be skipped, and folders with completed *hf.out* files are not regenerated unnecessarily. The scope of this stage is limited to job creation and optional submission. It does not attempt to diagnose quantum chemistry failures, license issues, module-loading problems, or convergence errors; those remain part of interpreting the completed job logs.

### Stage 3: Training-Set Data Generation

The third stage converts completed quantum chemistry outputs into ReaxFF training data. The current implementation supports parsing Jaguar *hf.out* files. For each completed isomer job, the workflow extracts the final total energy and optimized geometry, then energies are converted from hartree to kcal/mol. The structures are subsequently sorted according to their calculated energies, and the lowest-energy isomer is selected as the reference structure. Then, relative energies are computed for the remaining structures, and finally these relative energies are written to the ENERGY section of the ReaxFF *trainset.in* file.

The primary ReaxFF training data files generated by the workflow are the *geo* and *trainset.in* files. Auxiliary *composition.txt* and log files are also generated to support composition tracking, validation, and downstream training preparation. The generated *trainset.in* file is compatible with ReaxKit's existing trainset parser, enabling the resulting training data to be inspected and further processed using other ReaxKit tools. This establishes a direct link between quantum-derived reference data and the inputs required for ReaxFF force-field optimization.

### Demonstration Results

Figure 6 shows an example of the structure-generation and energy-evaluation stages of the workflow. **Specifically,** Figure 6-A shows the initial $C_5H_{13}B_3$ structure together with several isomeric configurations identified during the ReaxFF MD simulation. The variety of molecular geometries illustrates the structurally diverse configurations that can be automatically extracted and subsequently evaluated using quantum-chemistry calculations. Figure 6-B provides a representative comparison of the relative energies predicted by ReaxFF and calculated using DFT for the $C_5H_{13}B_3$ structures generated at 2500 K. ReaxFF generally reproduces the overall increase in relative energy across the ordered structures, although larger deviations and fluctuations are observed for some of the higher-energy configurations. This comparison illustrates how the automatically generated quantum-mechanical reference energies can be used to assess the ReaxFF predictions and provide training targets for force-field optimization.

A

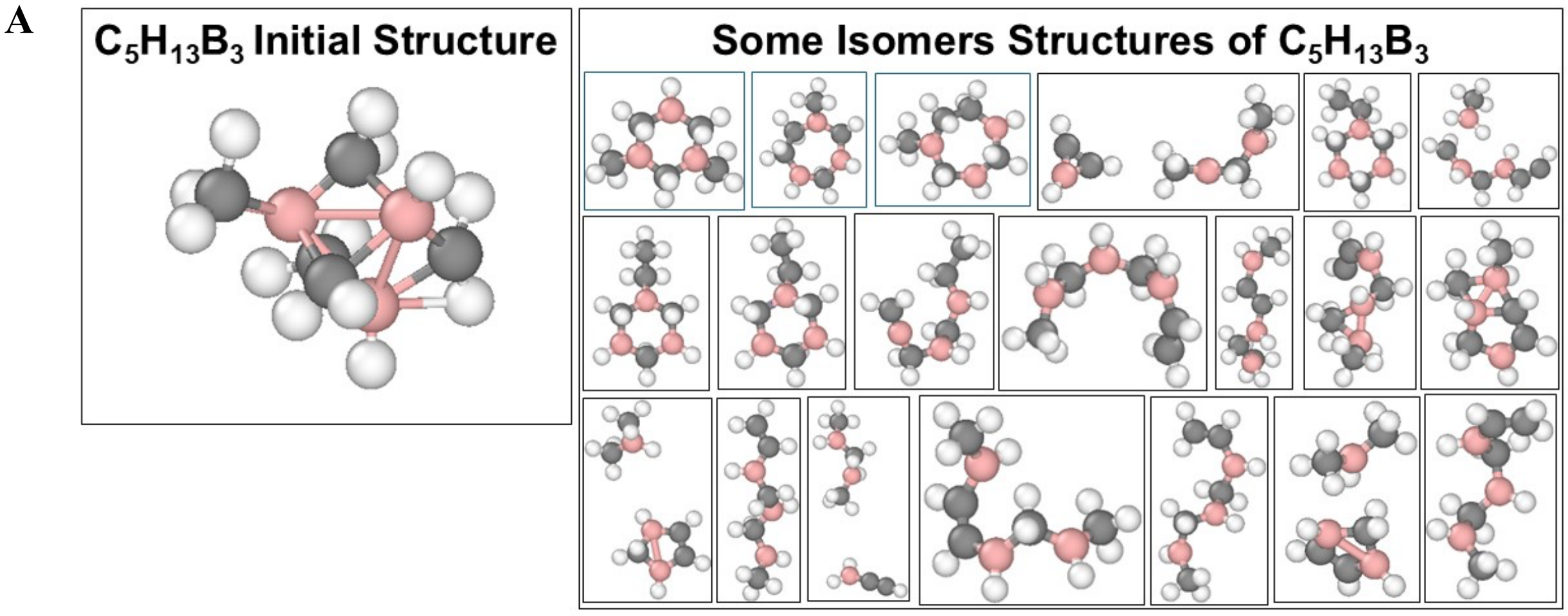

B

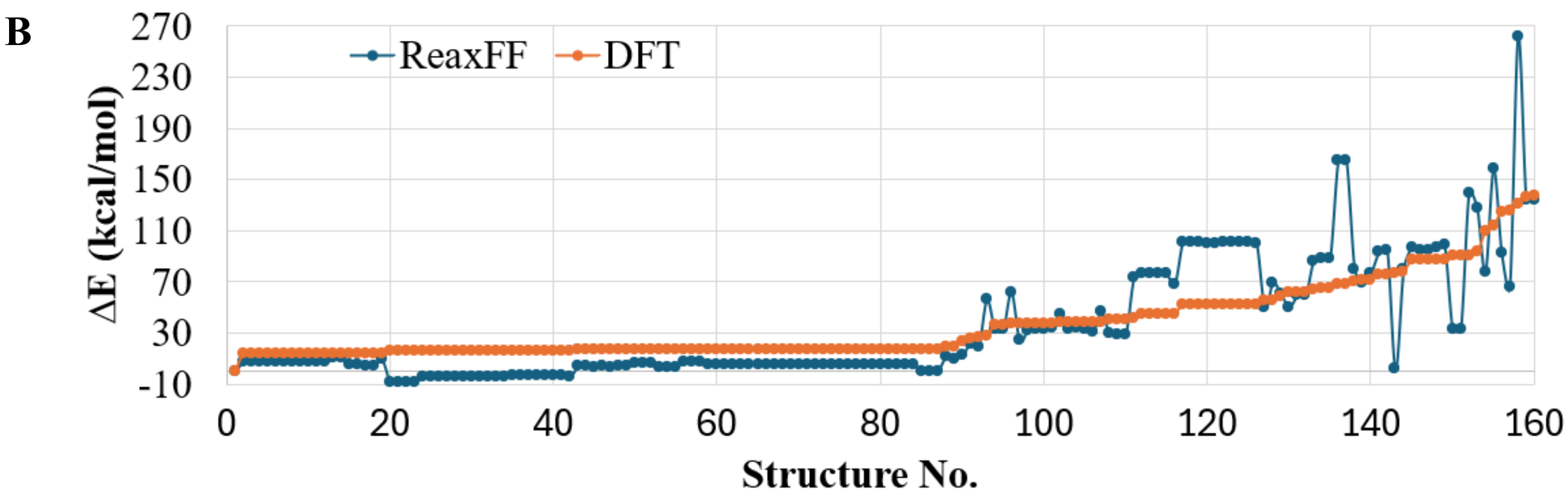

**Figure 6: ReaxFF-generated $C_5H_{13}B_3$ isomers and energy comparison.** (A) Initial $C_5H_{13}B_3$ structure and representative isomers identified from the ReaxFF MD trajectory for subsequent quantum-chemistry calculations. (B) Comparison of the relative energies (ΔE) predicted by ReaxFF and calculated using DFT for structures obtained from the 2500 K ReaxFF MD simulation.

## 3.2. Force-Field Merging

ReaxFF force-field development frequently requires combining parameters from different force-field files, for example when extending an existing force field to include additional elements or interactions. Performing this operation manually can be cumbersome because ReaxFF parameter files identify atoms and their interactions numerically according to the atom-type indices defined within each force field. For example, a bond entry beginning with *2 2* represents an interaction between two atoms assigned atom-type index 2, rather than explicitly identifying the corresponding element. Because these indices can differ between force-field files, interpreting and transferring parameters requires repeatedly referring back to the atom-type definitions and carefully renumbering interaction entries. This process becomes particularly time-consuming and error-prone when many bond, angle, torsion, off-diagonal, or hydrogen-bond parameters must be transferred.

To simplify this process, ReaxKit provides the merge-ffield command, which merges selected atom types and their associated interactions from a *source* force field into a *destination* force field. The user specifies the elements to transfer using *--atom-types* and can optionally restrict the operation to particular parameter sections, including *atom*, *bond*, *off_diagonal*, *angle*, *torsion*, and *hbond*. ReaxKit automatically maps the source atom indices to the corresponding destination indices and transfers compatible interactions involving the selected elements. Existing destination parameters are preserved by default, while *--replace-existing* can be used when matching destination terms should instead be replaced by those from the source force field.

The command also provides options for cases in which the two force fields are not directly compatible. For example, *--keep-atoms-in-dest* can be used to reduce the destination force field before merging, while *--fill-missing-with-template* can generate missing interactions using a chemically similar atom already present in the destination force field. Internally, ReaxKit parses the force fields into structured parameter sections, performs the required atom-index mapping and interaction normalization, and determines whether each parameter should be appended, replaced, skipped, or generated from a template. The merged force field is written as a new file, leaving the original destination unchanged, and audit reports summarize appended, updated, skipped, incompatible, and template-generated parameters.

As a representative example, two force fields were merged using the command *reaxkit merge_ffield --source .\ffield2 --destination .\ffield1 --output merged_ffield --atom-types Al --fill-missing-with-template*. The destination force field contained C, H, O, N, S, Mg, P, Na, Ti, Cl, F, Ba, and B, while the source force field contained C, H, O, N, B, Al, and He. The objective was to introduce Al and its associated interaction parameters into the destination force field. Because *--fill-missing-with-template* was enabled, ReaxKit also attempted to generate any missing Al-containing interactions using the most chemically similar atom already present in the destination force field. The command produced the merged force-field file together with a *merged_merge_audit* directory that records how the merge was performed.

The audit directory separates the different stages of the operation into several folders. *source_snapshot* contains the original source parameters selected for transfer before their atom indices were remapped, whereas *destination_projection* shows how those parameters appear after conversion to the atom-indexing scheme of the merged destination force field. *skipped_existing* records candidate interactions that were not added because equivalent entries already existed in the destination, which is particularly useful for verifying that duplicate terms were avoided when *--replace-existing* was not enabled. Each of these directories contains section-specific records for *atom*, *bond*, *off-diagonal*, *angle*, *torsion*, and *hydrogen-bond* parameters, along with a *snapshot_summary.txt* file, allowing the user to inspect the merge at the individual parameter level.

Because some Al-containing interactions were not available directly from the source force field, ReaxKit also generated missing terms from a template atom. The *template_snapshot* directory records the destination entries used for this purpose. In this example, B was selected as the template for Al. The accompanying *similarity_summary.txt* documents this selection process and reports the evaluated candidates using descriptors such as chemical family, periodic-table group, and atomic-radius difference. Together, these audit outputs make the merge traceable: the user can inspect which parameters originated from the source, how their numerical atom indices were translated into the destination force field, which terms were skipped as duplicates, and which missing interactions were generated from a template.

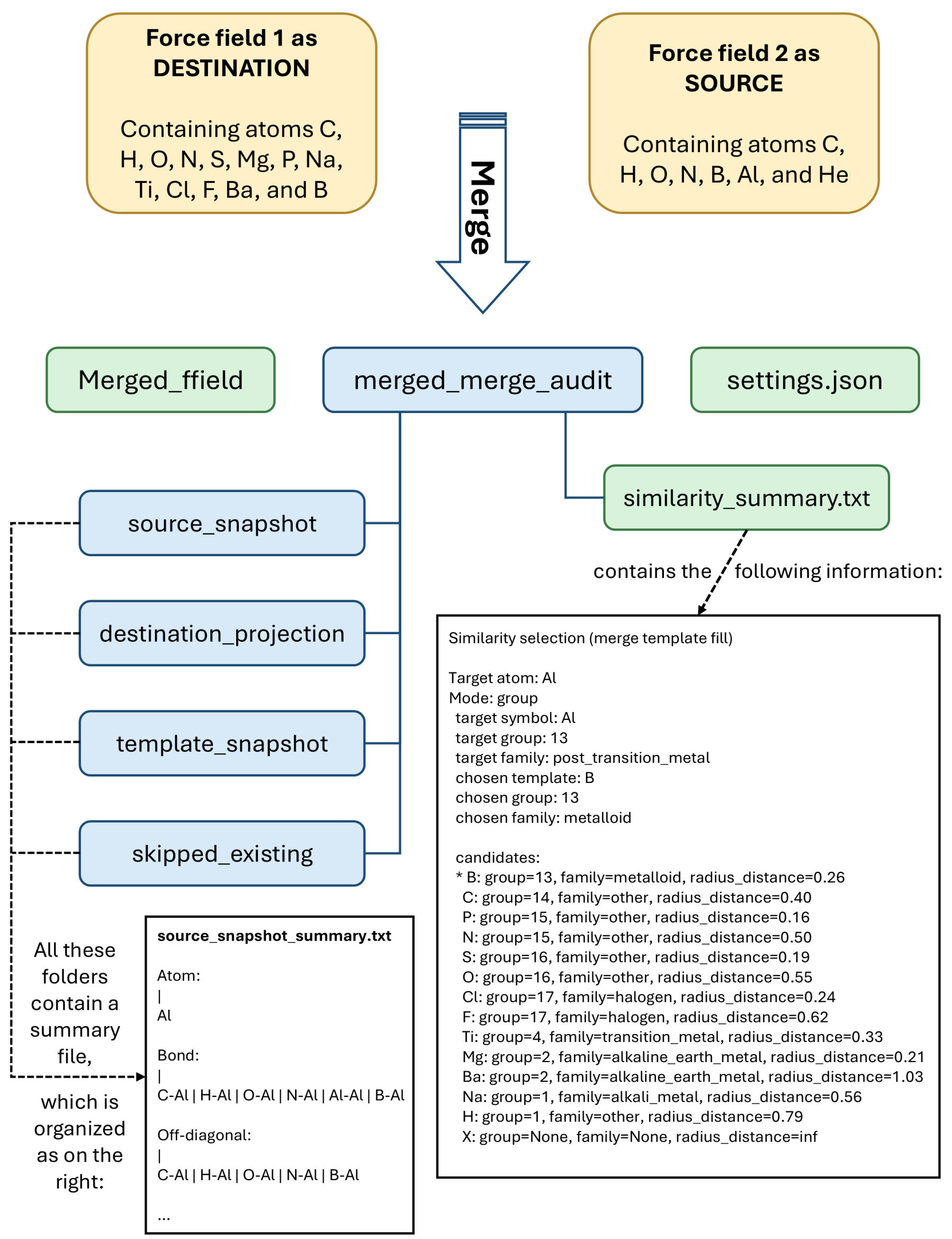

**Figure 7: Example ReaxKit force-field merging workflow and audit output.** Force field 1 is used as the destination and force field 2 as the source to introduce Al-containing parameters into the destination. The workflow generates the merged force field together with audit information that documents transferred, mapped, skipped, and template-generated parameters, including the similarity-based selection of a template atom for missing Al interactions.

## 3.3. Force-Field Optimization Diagnostics

During ReaxFF force-field optimization, sequential parabolic optimization varies individual parameters within prescribed ranges and evaluates the resulting objective-function error. The following sections present two complementary diagnostics for examining ReaxFF force-field optimization: sensitivity analysis identifies which parameters most strongly affect the force-field error, while parameter-evolution analysis shows how those parameters were explored during optimization and where their final values converged.

### 3.3.1. Force-field optimization sensitivity analysis

The force-field optimization diagnostics file, *fort.79*, records the parameter values sampled during optimization together with the corresponding objective-function responses. ReaxKit uses these data through the *get_ffield_diagnostics_sensitivity* command to quantify how sensitive the force-field error is to changes in individual parameters. For each parameter, the sensitivity (S) is evaluated from relative objective-response ratios, as shown by Eq. (1):

$$\boldsymbol{S_{\pm}} = \frac{\boldsymbol{E(p \pm \Delta p)}}{\boldsymbol{E(p)}} \tag{5}$$

where $E(p)$ is the objective-function value at the current parameter value $p$, and $E(p \pm \Delta p)$ is the objective-function value after perturbing that parameter by $\pm\Delta p$. The resulting sensitivities can be visualized using either a beeswarm or tornado representation. For example, the command *reaxkit get_ffield_diagnostics_sensitivity --plot beeswarm* generates a beeswarm plot in which each row corresponds to a force-field parameter and each point represents one sampled objective response, as shown in **Figure 8**. The point color indicates the corresponding parameter value, ranging from lower values in blue to higher values in red. It is worth noting that flags such as *--interpret* can replace numerical parameter identifiers with descriptive force-field labels, and *--vline 1.0* can add a reference line for interpreting the relative response.

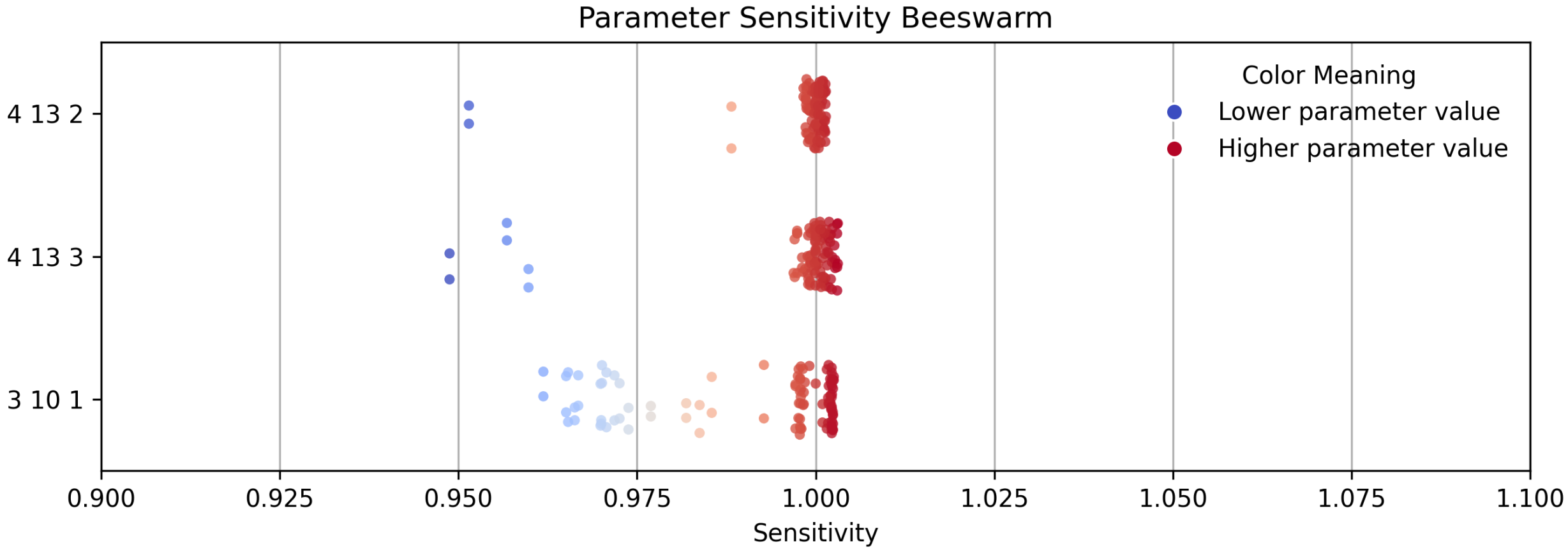

**Figure 8: Force-field parameter sensitivity obtained from ReaxFF optimization diagnostics.** Beeswarm representation of the relative objective response for selected force-field parameters. Each point represents a sampled parameter value, with color ranging from lower parameter values (blue) to higher values (red); a wider horizontal distribution indicates a stronger objective response to variation of that parameter.

In the representative beeswarm plot shown in **Figure 8**, the three displayed force-field parameters exhibit distinctly different response distributions. Parameter 3 10 1 spans the widest sensitivity range, with sampled responses extending from approximately 0.96 to 1.00, indicating a stronger dependence of the optimization objective on this parameter over the explored range. In comparison, parameters 4 13 2 and 4 13 3 are concentrated much more closely around a sensitivity value of 1.0, suggesting a weaker response within the sampled region. Using sensitivity analysis, users can identifies the parameters that most strongly influenced the objective function, helping prioritize force-field refinement while revealing weakly constrained or low-impact parameters.

### 3.3.2. Evolution of force-field parameters during optimization

Sensitivity analysis describes how strongly the objective responds to individual parameters, but it does not by itself show how the optimizer explored those parameters or where the optimized values ultimately converged. For this purpose, ReaxKit provides the *get_ffield_diagnostics_evolution* command, which reconstructs the evolution of the force-field decision variables from the optimization diagnostics. Examining this evolution is useful for determining whether parameters were explored broadly or only within a narrow region, whether the final values moved substantially away from their initial values, and whether optimization repeatedly sampled values close to the imposed parameter bounds. The command combines information from three files: a force field optimization diagnostic file (i.e., *fort.79*), *ffield*, and *params*. Each sampled parameter value is normalized according to Eq. (1):

$$\boldsymbol{x}_{\textbf{norm}} = \frac{\boldsymbol{x} - \boldsymbol{x}_{\textbf{lower}}}{\boldsymbol{x}_{\textbf{upper}} - \boldsymbol{x}_{\textbf{lower}}}, \tag{6}$$

so that values of 0 and 1 correspond to the lower and upper optimization bounds, respectively.

**Figure 9** has been obtained using *reaxkit get_ffield_diagnostics_evolution --save parameter_evolution.png* command on a sample data, where each row represents one optimized force-field parameter, while the horizontal position of each point indicates the normalized parameter value sampled during optimization. The color of each point represents the corresponding objective-function value, enabling the sampled regions that produced relatively low or high optimization errors to be identified. The information displayed to the right of each row additionally reports the parameter's optimization bounds, starting value, and final optimized value. This makes it possible to assess not only where the optimizer sampled a parameter but also how far the final solution moved from its initial value.

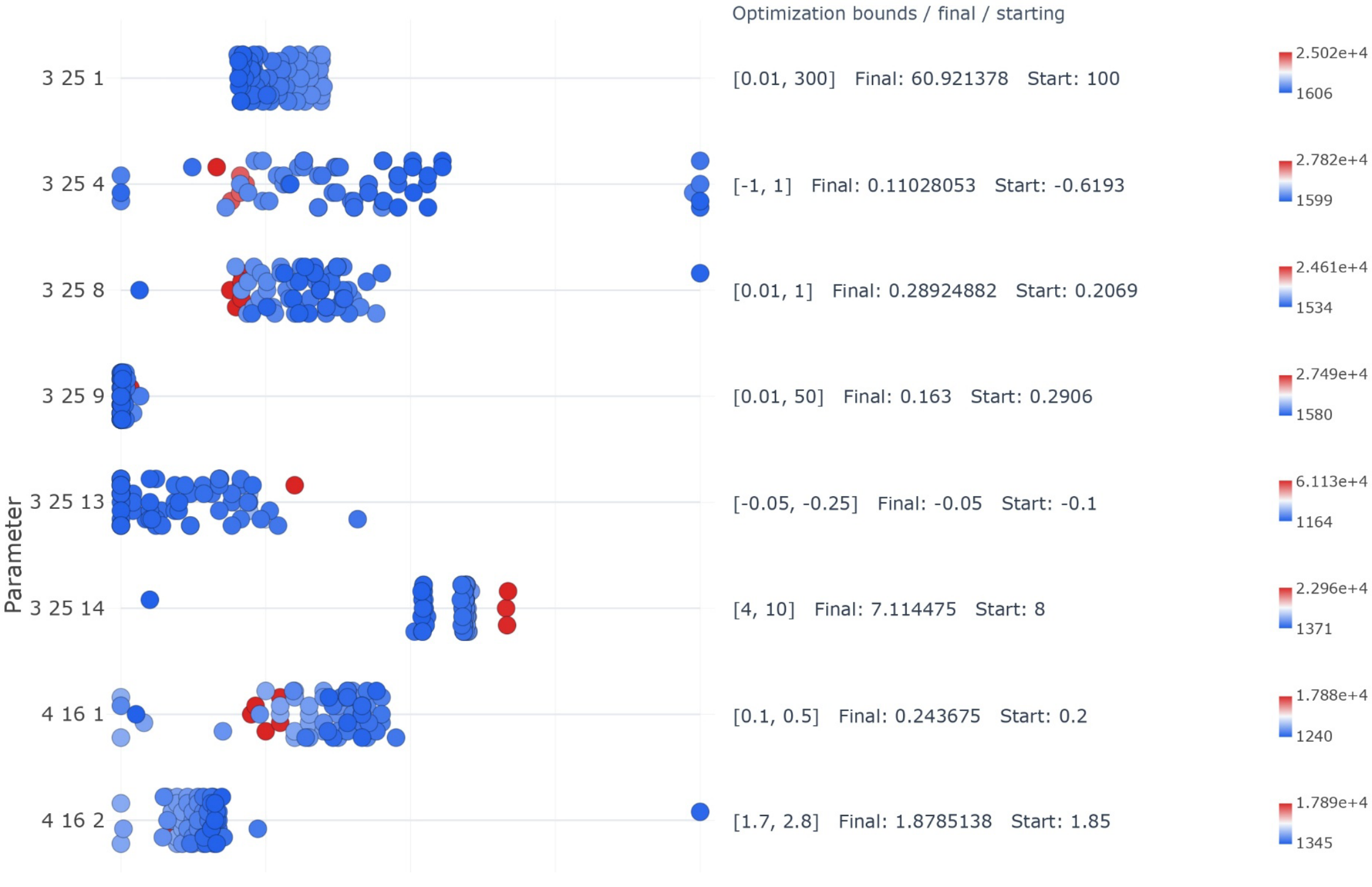

**Figure 9: Evolution of force-field parameters during ReaxFF optimization.** Sampled values of individual force-field parameters are shown after normalization to their prescribed optimization bounds, where 0 and 1 correspond to the lower and upper bounds, respectively. Point color represents the associated objective-function value, while the optimization bounds and starting and final parameter values are reported for each parameter.

The figure reveals qualitatively different optimization behavior among the parameters. Some parameters, such as 3 25 1, were sampled within a relatively narrow portion of their allowed range, whereas others, such as 3 25 4 and 4 16 1, were explored over much broader normalized intervals. Several parameters also contain isolated sampled values far from the main cluster, showing that the optimizer evaluated more distant regions before returning to a narrower range. Comparison of the starting and final values further indicates whether optimization produced only a minor correction or substantially changed a parameter from its initial force-field value.

This information is particularly valuable for diagnosing optimization behavior. A parameter that repeatedly approaches one of its imposed bounds may indicate that the selected optimization interval should be reconsidered, whereas a parameter that remains near its starting value may already be close to a favorable region or may have little influence on the objective.

### 3.4. Active site analysis

Active-site analysis is included in ReaxKit as a dedicated analysis package for identifying and characterizing structurally or chemically important local environments in ReaxFF simulations. Unlike generic trajectory or property analysis, this capability focuses on site-specific information, including local classification, defect-related descriptors, periodic-boundary handling, and event-based changes around selected atoms or regions. The active-site module in ReaxKit was implemented to analyze where reactions occur within a reactive MD trajectory and what local structure those reacting atoms had before or during the event.

In the present implementation, this workflow is applied to graphitic carbon membranes exposed to oxygen- and silicon-containing species. The carbon framework is treated as the set of candidate active sites, while oxygen and silicon atoms are treated as reacting species whose persistent contacts with carbon are followed over time. The workflow has two parts. First, a structural task analyzes a selected frame and constructs a bond graph for the chosen substrate or network element. Second, an event task samples the trajectory and records persistent C-O and C-Si contacts using either ReaxFF bond orders or distance-based criteria. In distance mode, ReaxKit can first run a diagnostic step to sample C-X distance distributions and contact lifetimes, so that the final cutoff and persistence threshold are chosen from the trajectory rather than assumed a priori.

For the carbon system considered here, graphitic descriptors are assigned only to carbon atoms. Gas-phase oxygen and silicon atoms are not assigned grain identity, edge type, ring membership, or pyramidalization values, because these quantities are defined for atoms belonging to the connected carbon network. Those gas-phase species instead enter the event table, where ReaxKit records whether each carbon atom forms a persistent C-O or C-Si contact, the first frame where the event is confirmed, and the number of confirmed events.

The structural table currently contains the following substrate-centered descriptors: total coordination, carbon-network coordination, under-coordination, heteroatom-bond flag, pyramidalization distance, angular deviation for two-coordinate atoms, bond strain, angle strain, local roughness, bond-orientational order, local orientation-domain ID, primitive-ring size range, non-hexagonal ring membership, defect type, boundary flag, edge label, boundary-segment ID, and final site label. Optional SOAP descriptors can also be computed for the selected substrate atoms when a higher-dimensional local-environment fingerprint is required. Periodic boundary conditions are handled during bond, distance, and local-geometry calculations.

Although the present example is carbon oxidation and carbon-silicon bonding on a graphitic membrane, the same analysis pattern can be extended to related systems, including other reactions on carbon, carbon deposition or etching, and catalytic surfaces where local atomic structure controls probable active sites. In such cases, the substrate element, reacting species, structural descriptors, and event criteria can be adjusted while retaining the same basic workflow: assign local structure, track reactive events, and merge the two tables by atom ID.

The active-site table can also guide more expensive follow-up calculations. If a large ReaxFF trajectory contains thousands of atoms, selected active sites can be used as centers for extracting local clusters within a chosen cutoff radius. After appropriate passivation of broken bonds, such clusters could be used for DFT calculations, force-field training, or validation data. This use is conceptually related to cluster-carving approaches such as *carve.f90 (Caro et al., 2014, 2020)*, which extract local atomic environments from large structures when the full system is too large for the electronic-structure method. In the present ReaxKit implementation, cluster extraction and passivation are not part of the active-site workflow; they are downstream operations enabled by the atom-resolved active-site output.

**Demonstration Results**

The active-site workflow was tested on a ReaxFF trajectory containing a reconstructed carbon structure and oxygen-containing gas phase. The purpose of this example is to show how ReaxKit can connect local carbon structure with later reactive events. The structural plots therefore include only carbon atoms. Oxygen atoms are present in the simulation and are used for event detection, but they are excluded from the structural maps because graphitic quantities such as grain identity, edge label, and pyramidalization do not have a meaningful interpretation for isolated gas-phase atoms.

Figure 10 shows the carbon-site labels assigned from the local bonding environment. Most atoms are classified as either basal-like or defect-associated, with a small number of under-coordinated edge atoms. For this frame, the carbon-only label counts are 1045 basal atoms, 744 defect-associated atoms, 4 armchair-edge atoms, and 7 zigzag-edge atoms. These labels should be read as structural categories for the carbon network, not as chemical identities of all atoms in the simulation box.

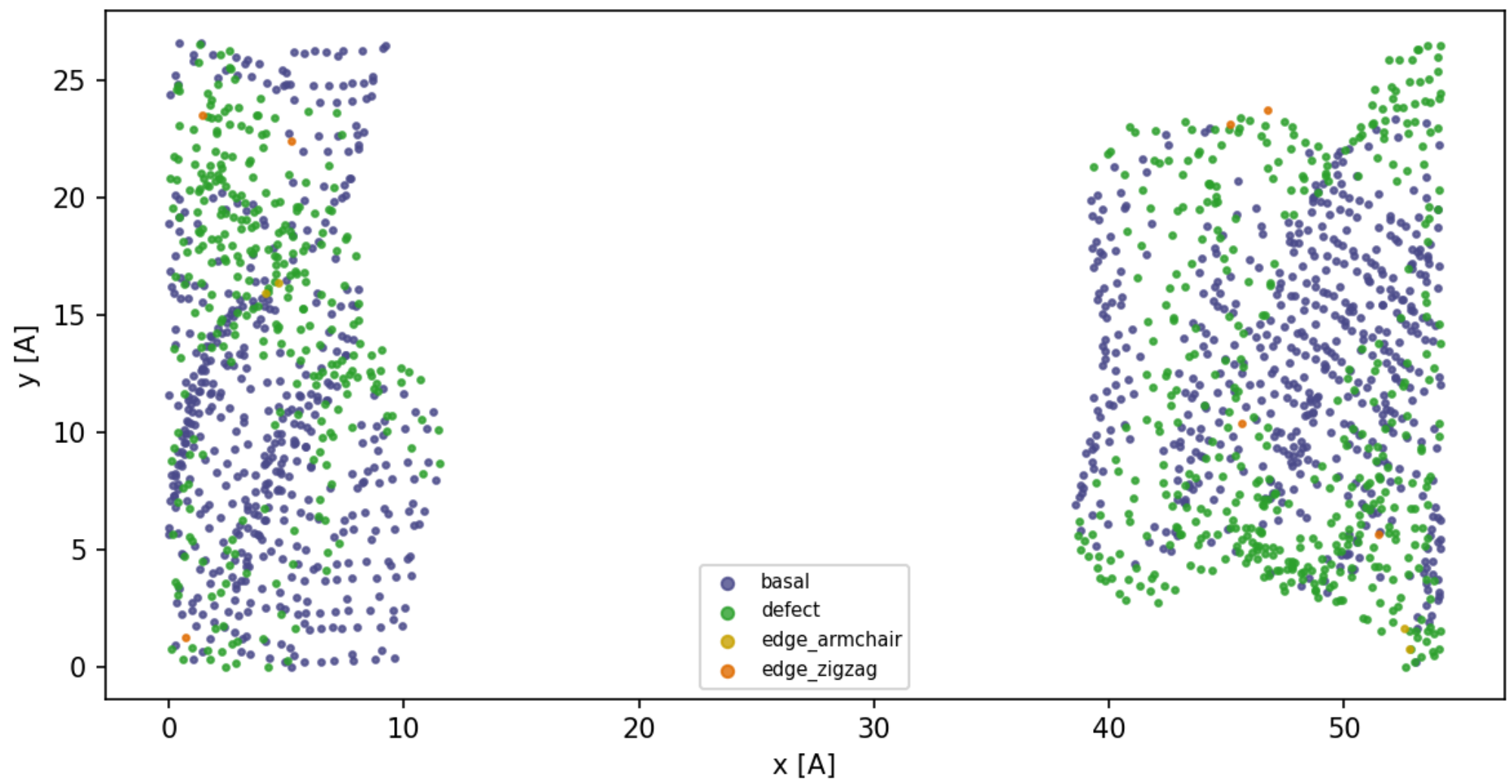

**Figure 10: Carbon-site labels from the active-site structural analysis.** Only carbon atoms are shown. Oxygen atoms in the gas phase are excluded from this structural map because graphitic labels such as basal, defect, zigzag edge, and armchair edge are defined for the connected carbon network.

Figure S5 shows a local orientation-domain map obtained from the bond-orientational order of the carbon network. The map indicates a highly fragmented reconstructed structure, with 420 assigned carbon orientation domains in this frame. This number is a descriptor of local orientational coherence in the analyzed frame. It is not an experimental crystallite-size measurement and should not be compared directly with La or Lc values from X-ray diffraction or microscopy. Figure S6 maps the absolute pyramidalization distance, |d_pyr|, for carbon atoms. Pyramidalization measures the out-of-plane displacement of a carbon atom relative to the plane defined by its bonded carbon neighbors, and therefore provides a local measure of $sp^2$-to-$sp^3$-like distortion. In this frame, the mean |d_pyr| over carbon atoms is 0.142 Å, the median is 0.112 Å, and 21.7% of carbon atoms exceed the 0.229 Å reference value used for this example. The red markers identify under-coordinated carbon atoms, which are treated separately because their reactivity is controlled primarily by missing coordination rather than by a three-neighbor pyramidalization measure. Finally, Figure S7 compares the |d_pyr| distributions for basal and defect-associated carbon atoms. Defect-associated atoms show a broader distribution and a stronger high-|d_pyr| tail, while basal atoms are concentrated closer to the planar limit. This supports the intended use of the workflow:

structural descriptors from one or more frames can be combined with event statistics from the trajectory to determine which local environments are most likely to participate in oxidation, carbon-silicon bonding, or related reactions.

### 3.5. Regeneration of Simulation Files Based on Derived Atomic Properties

Another representative capability of ReaxKit is its ability to regenerate simulation files, such as the trajectory file, using properties derived from analysis. This is useful when the quantity of interest is not the chemical identity of an atom itself, but its evolving local environment during a simulation. For example, coordination changes can provide direct information about bond breaking and formation during processes such as surface bombardment, polymer-chain growth, deposition, or other reactive events. ReaxKit can therefore combine information from multiple simulation files, analyze the local state of each atom, and write a new trajectory in which the atom labels encode the resulting structural classification.

This workflow is implemented through the *relabel_traj_using_coordination* command. For example, *reaxkit relabel_traj_using_coordination --valences 'Al=3,N=3,He=0' --output xmolout_relabeled --mode global --labels=-1=U,0=C,1=O* reads atomic coordinates together with bond-order and connectivity information. For every atom and frame, ReaxKit determines its coordination from the available bonding information and compares it with a user-defined target coordination, such as 3 for Al and N and 0 for He. Based on this comparison and a specified tolerance, each atom is classified as under-coordinated, normally coordinated, or over-coordinated.

The resulting coordination status can then be used to replace the original chemical labels in the trajectory. With *--mode global*, atoms are labeled according to their coordination state, for example U, C, and O for under-coordinated, coordinated, and over-coordinated atoms, respectively. Alternatively, *--mode by_type* can retain both chemical identity and coordination information through labels such as AlU, AlC, or AlO for Al atoms. ReaxKit then writes the relabeled trajectory to a new trajectory file while preserving the atomic coordinates and frame structure. The underlying coordination data can also be exported separately, for example using *--export coordination.csv*.

The regenerated trajectory can subsequently be opened directly in visualization tools such as OVITO (Stukowski, 2010), where the coordination-based labels can be assigned different colors or visual styles. This makes it possible to follow the spatial and temporal evolution of coordination changes atom by atom, rather than reconstructing them manually from separate trajectory and connectivity files. **Figure 11** shows an example of a single frame for a coordination-based trajectory rendered by OVITO (Stukowski, 2010). This trajectory is for a He-irradiated AlN slab, and is obtained by running the command *reaxkit relabel_traj_using_coordination --valences 'Al=3,N=3,He=0' --output xmolout_relabeled --mode global --labels=-1=U,0=C,1=O*.

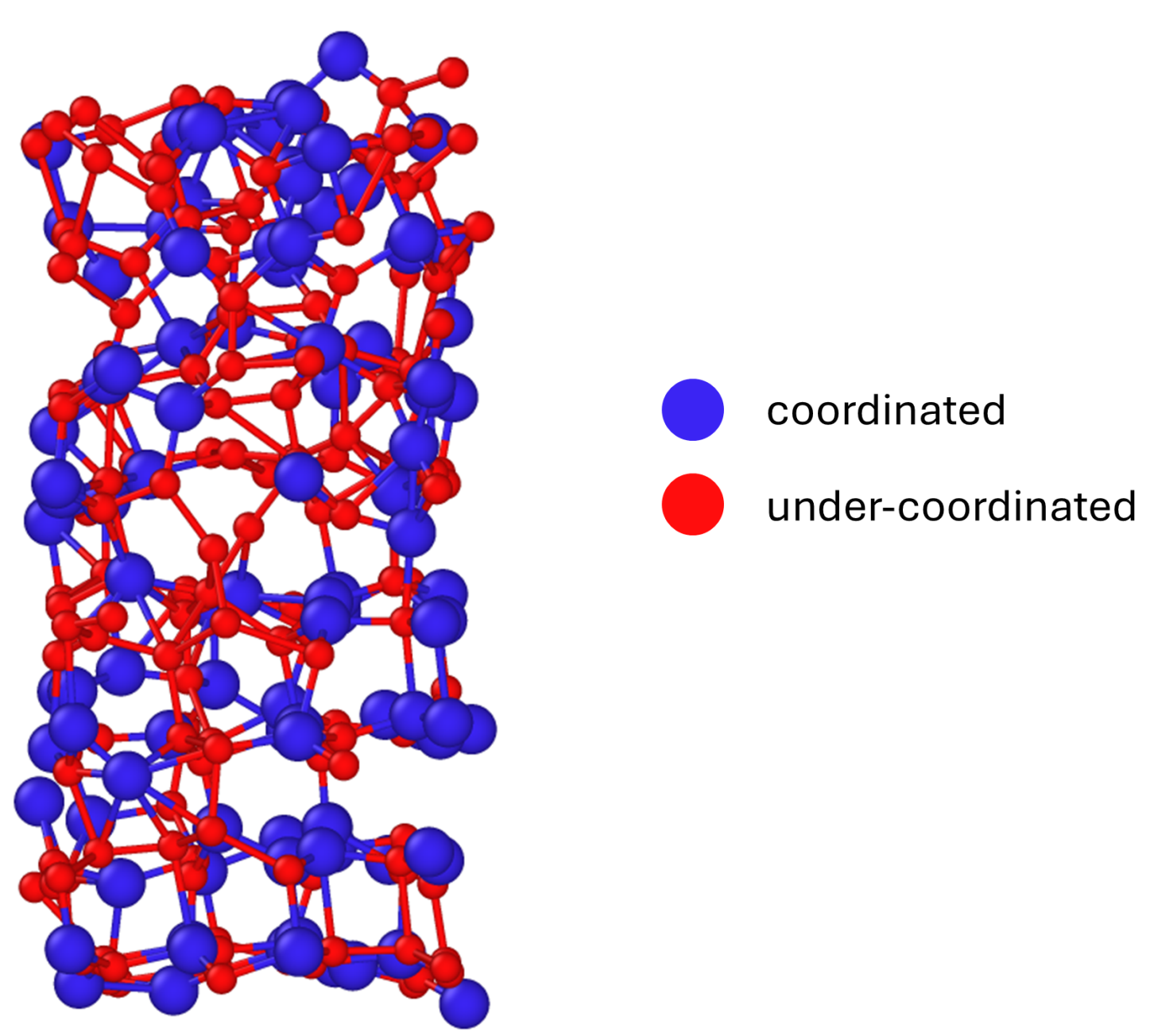

**Figure 11: Example visualization of a ReaxKit coordination-relabeled trajectory.** Atoms are classified according to their coordination state, with coordinated, under-coordinated, and over-coordinated atoms shown in blue, red, and yellow, respectively, enabling direct visualization of local coordination changes during the simulation.

### 3.6. Study design

The study-design workflow in ReaxKit provides an end-to-end framework for converting a research plan into a reproducible multi-run simulation campaign. Instead of manually creating folders, editing repeated input files, launching individual simulations, and tracking outputs across many cases, the user defines the study, runs the simulations, performs the analysis, aggregates data, and presents them as appropriate. This workflow combines ReaxKit's input-generation and output-processing capabilities into a single automated pipeline.

**Figure 12** shows an example study in which the user varies two parameters, magnesium concentration and temperature, with two replicates for each simulation case. Each case consists of a three-stage simulation sequence: energy minimization (MM), NPT equilibration, and NVT production. The user first creates a template folder containing a representative MM → NPT → NVT simulation cascade. Next, the user generates the template YAML file which defines the study settings, and modifies it according to his desired settings. ReaxKit then uses this information to initialize the study folder structure, generate or organize the required input files, execute the planned stages, track status files and manifests, and preserve the relationship between each case, replicate, and output. After the simulations are completed, the user can run group-level analysis, aggregate results across all cases and replicates, and generate final plots or tables for interpretation.

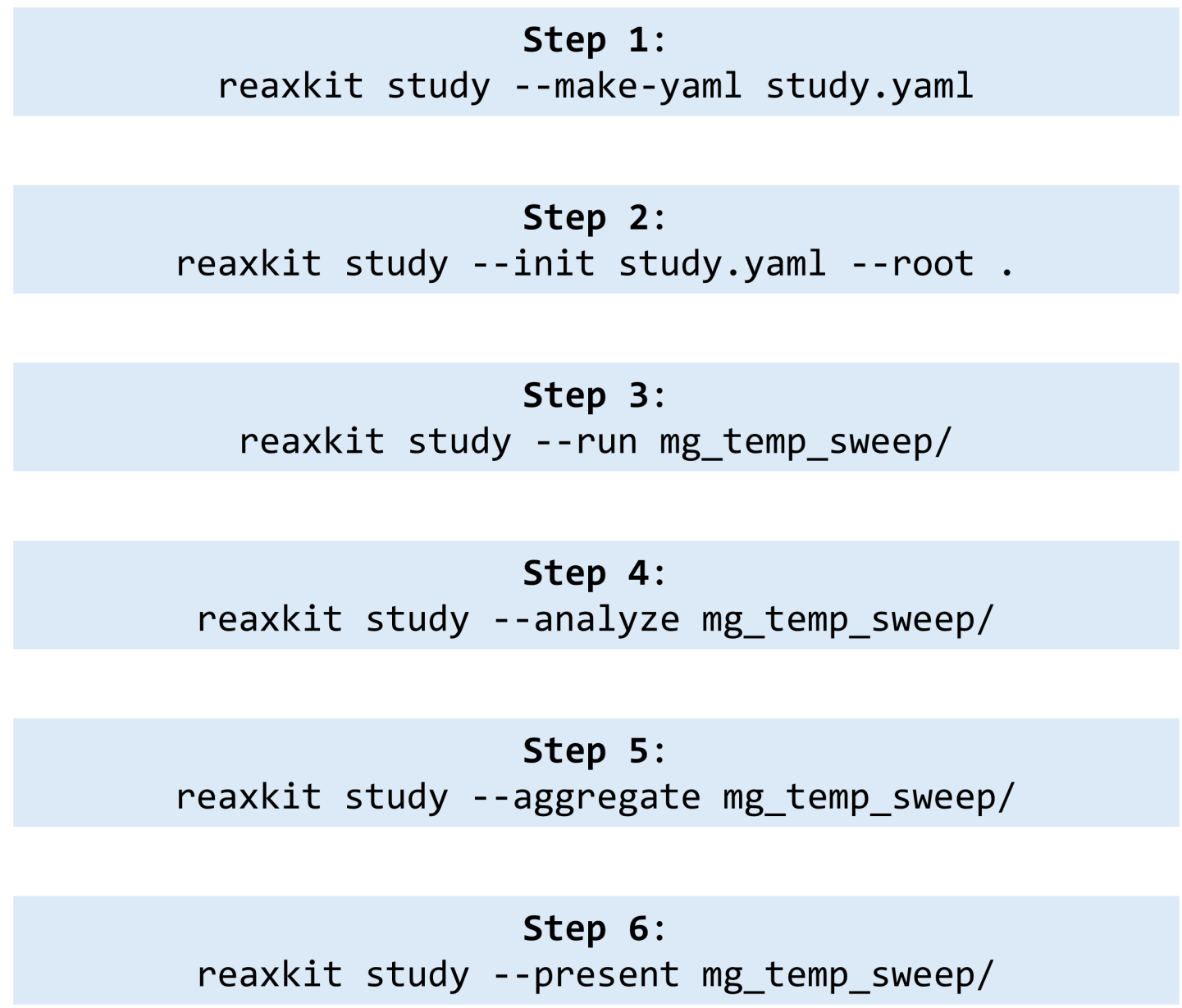

**Figure 12: Stepwise command-line workflow for ReaxKit study design.** The study-design workflow begins by creating a YAML study template, followed by initializing the study folder structure, running the planned simulation stages, analyzing individual outputs, aggregating results across cases and replicates, and generating final presentation outputs. Together, these commands illustrate how ReaxKit turns a user-defined research plan into a structured, reproducible, and automatable multi-run simulation study.

Figure S8 shows how the template YAML file is modified for the Mg–temperature sweep example. The first section defines the overall study configuration, including the study name, template folder, simulation parameters (mg and temperature), and the number of replicates. The run section then defines the ordered simulation stages, including MM, NPT, and NVT. Within each stage, the user can specify one or more execution steps, such as geometry generation, control-file modification, job submission, or post-processing commands. For example, the command *python hybrid_generator.py Z {mg_percent}* calls a user-provided Python script that adds magnesium atoms to a zinc oxide slab. By writing *{mg_percent}* as a placeholder, the user allows the study workflow to automatically substitute each value defined in the parameter list, such as *mg: [40, 60]*, and generate the corresponding simulation cases. Similarly, the command *sbatch submit_and_wait.sh* submits each stage as a SLURM job on an HPC cluster and waits for that job to finish before the workflow proceeds to the next stage. The YAML file also records produced and consumed artifacts, allowing outputs from one stage, such as the final geometry from MM, to be passed automatically as inputs to a later stage, such as NPT. After the simulation stages, the analysis section defines which analysis should be performed on completed runs, which stage should be analyzed, and which output variables should be collected. Finally, the aggregate section specifies how analysis results should be combined across cases and replicates, including the selected x and y variables, statistical summaries, and missing-data behavior.

This structure also provides a foundation for future autonomous simulation campaigns, where decision rules or AI agents could update the YAML configuration, select the next parameter set, launch new simulations, and analyze results iteratively with minimal manual intervention.

**Demonstration Results**

Once the study is initialized and run, the folder organization shown in **Figure 13** is created under the *mg_temp_sweep* study folder. The *cases/* directory contains one folder for each parameter combination, resulting in four cases for the combinations of *mg_percent = 40 or 60* and *temperature = 300 or 500*. Each case folder contains two replicate folders, and each replicate follows the same staged simulation sequence, including MM, NPT, and NVT. At the top level, *study_manifest.json* records the complete study configuration and provenance, including the source YAML file, template directory, parameter values, case identifiers, replicate paths, run stages, analysis definitions, and the total number of generated runs. The *study_run_status.csv* file tracks the execution status of the different runs and stages, allowing the user to monitor progress and identify incomplete or failed steps. Similar status and tracking reports are also generated when the user runs the analysis and aggregation steps, so simulation execution, post-processing, and summary generation remain traceable across all cases and replicates.

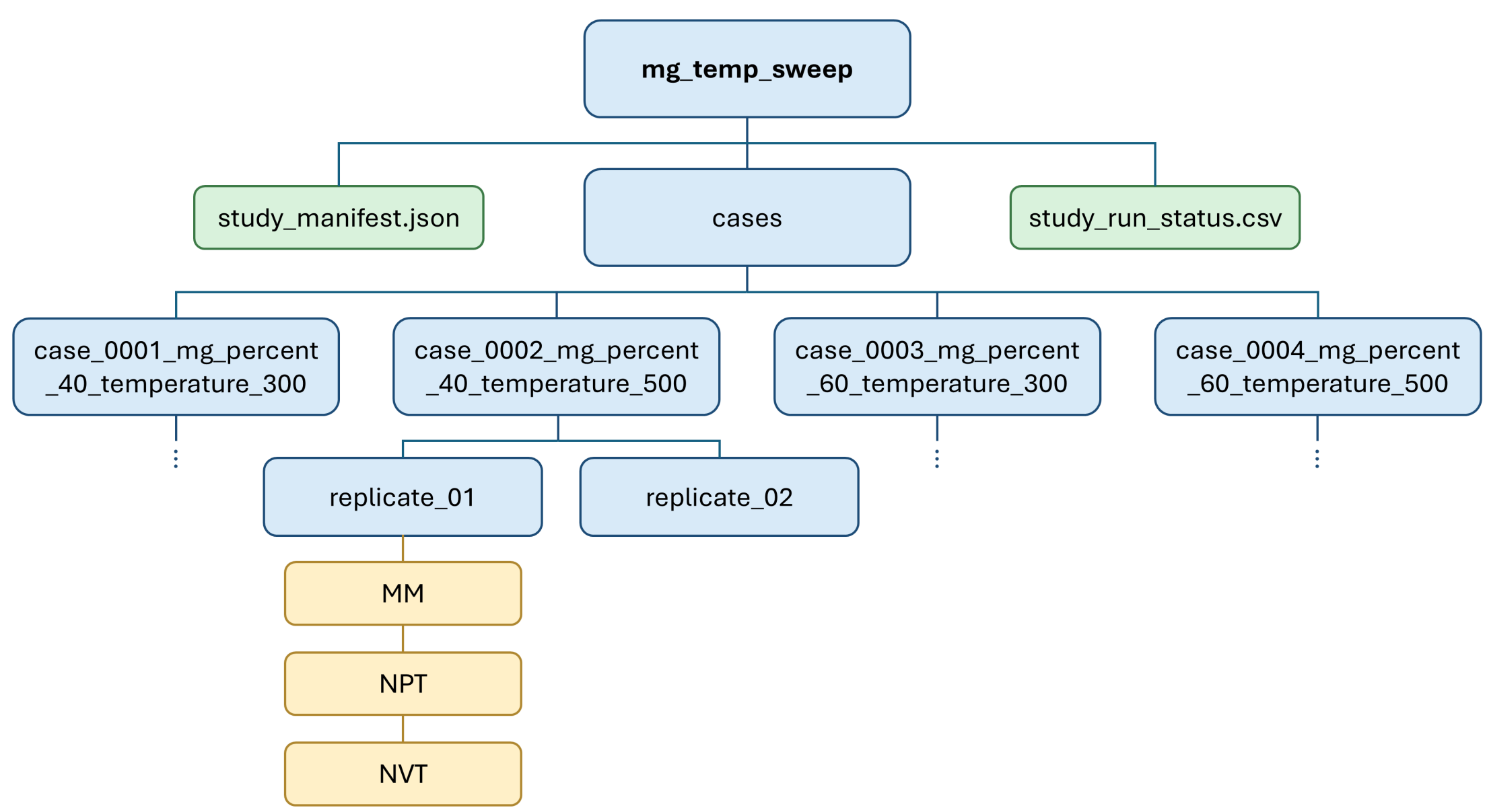

**Figure 13: Folder organization generated by the ReaxKit study-design workflow.** The *mg_temp_sweep* study folder contains a manifest file, a run-status table, and a *cases/* directory with one case folder for each magnesium concentration–temperature combination. Each case contains replicate folders, and each replicate follows the same MM → NPT → NVT simulation sequence. This structure preserves the relationship between study parameters, cases, replicates, simulation stages, and generated outputs, enabling reproducible execution, analysis, aggregation, and reporting across the full simulation campaign.

# 4. Conclusion

This work introduced ReaxKit, a modular Python toolkit developed to reduce the manual effort associated with preparing, parsing, organizing, and analyzing ReaxFF molecular dynamics simulations. Its primary contribution is the creation of a reusable workflow layer between ReaxFF simulation engines and scientific analysis, addressing a gap not fully covered by existing tools that focus on specialized tasks such as reaction-network analysis or force-field optimization. By converting engine-specific files into consistent internal representations and exposing common operations through unified interfaces, ReaxKit lowers the technical barrier to routine ReaxFF use, improves reproducibility, and supports the transition from isolated, manually processed simulations toward scalable and high-throughput computational studies.

ReaxKit was developed using a separation-of-concerns architecture in which engine-specific file handling, canonical domain models, scientific analysis, workflow orchestration, presentation, storage, and graphical interaction are implemented as distinct but interoperable layers. Engine adapters normalize outputs from Standalone ReaxFF, LAMMPS, and AMS into typed data classes, while analyzers follow a consistent Request–Task–Result pattern. Registered workflows and the *AnalysisExecutor* then connect user requests to the required data, execute the corresponding analysis, and dispatch structured results to plotting, reporting, or export backends. The accompanying *reaxkit_workspace*, caching system, provenance identifiers, command-line interface, help system, and local browser-based graphical interface further support traceable, maintainable, and accessible execution across personal workstations and high-performance computing environments.

The representative applications demonstrated that ReaxKit extends beyond basic file conversion. The elastic training-set generator retrieved structural and mechanical information from the Materials Project and converted it into strained geometries, energy–strain and energy–volume data, ReaxFF training-set entries, and diagnostic status reports. The active-site analysis modules identified and characterized local structural environments, defects, grains, under-coordinated atoms, and pyramidalization behavior. In addition, the study-design workflow translated a YAML-defined research plan into a structured multi-parameter and multi-replicate simulation campaign, including staged simulation execution, artifact transfer, status tracking, analysis, aggregation, and presentation. Together, these examples showed how the same architecture can support input generation, scientific interpretation, and campaign-level automation within a unified environment.

ReaxKit therefore provides both an immediately useful collection of ReaxFF utilities and an extensible foundation for future computational materials workflows. Additional engine adapters, analyzers, generators, renderers, and reporting tools can be incorporated without redesigning the underlying execution model. Future development of ReaxKit can focus on expanding support for additional ReaxFF engines and file formats, increasing the number of scientific analysis and input-generation modules, and establishing broader validation and benchmarking against established tools and reference datasets. Further work may also strengthen interoperability with external visualization, force-field optimization, materials databases, and high-performance computing platforms, while improving automated testing, documentation, and graphical workflows for users with limited programming experience. A particularly promising direction is the extension of the study-design framework toward adaptive and autonomous simulation campaigns, in which uncertainty-aware algorithms or AI agents select new simulation conditions, launch calculations, analyze the resulting data, and iteratively refine the next set of experiments. These developments could enable ReaxKit to evolve from a pre- and post-processing toolkit into a more comprehensive and intelligent workflow environment for reproducible, high-throughput reactive molecular dynamics research.

### Acknowledgments

AMD, AS and ACTvD acknowledge funding for ReaxKit conceptualization and development with support from the Center for 3D Ferroelectric Microelectronics (3DFeM), an Energy Frontier Research Center funded by the U.S. Department of Energy, Office of Science, Office of Basic Energy Sciences, under Award No. DE-SC0021118. MM and AUH acknowledge funding from AFOSR MURI # FA9550-24-1-0301 for isomer analysis tool development and transer to ReaxKit. JD and SMM acknowledge funding from ONR MURI # N000142412313 for developing analysis tools and ReaxKit debugging. AS and ACTvD also acknowledge funding from the Pennsylvania State University PCL-Test Bed LATTICE node under NSF cooperative agreement ITE-2607536.

### Data Availability

ReaxKit is publicly available through its GitHub repository at https://github.com/ali-m-dinani/reaxkit, with user documentation provided at https://ali-m-dinani.github.io/reaxkit/. The package is also distributed through PyPI and can be installed directly using *pip install reaxkit*, allowing users to easily install and update ReaxKit through standard Python package-management tools.

# Supporting Information

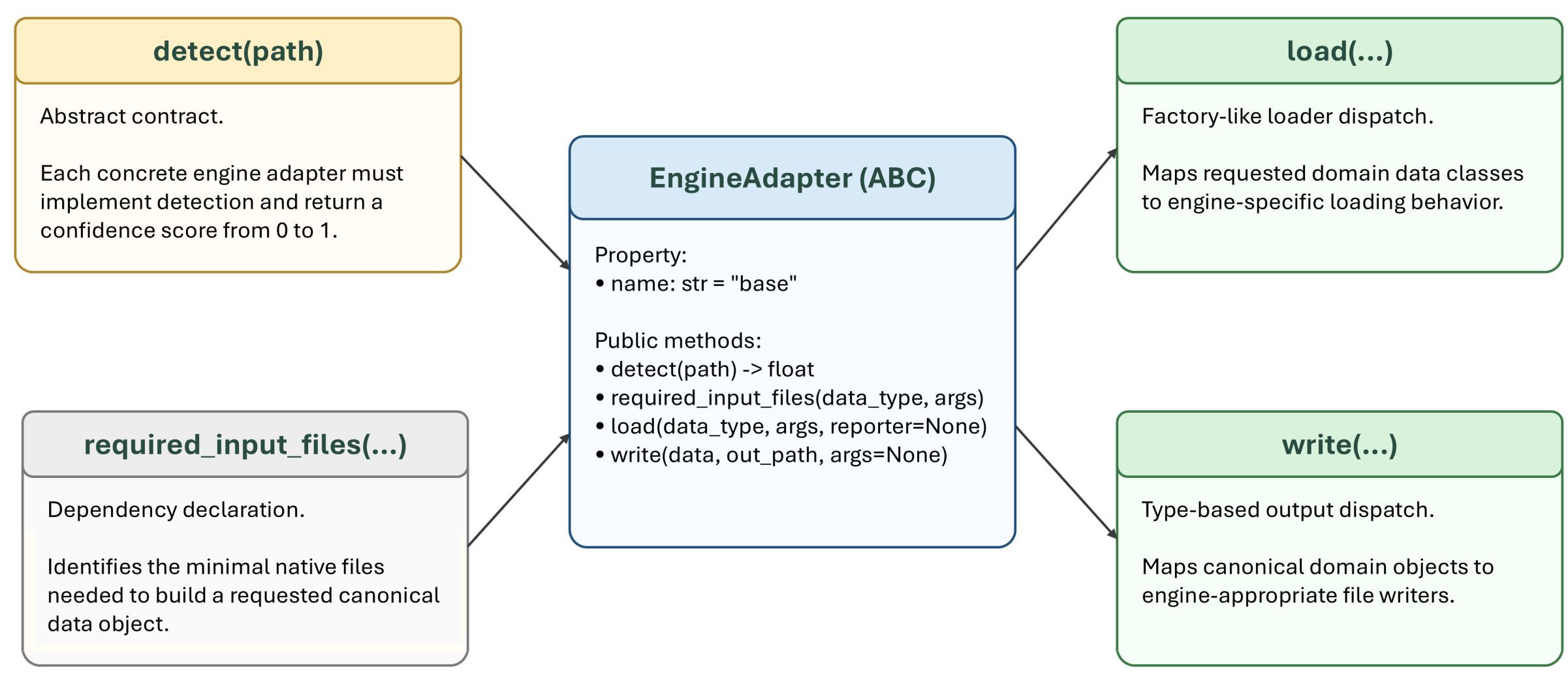

**Figure S1: Generic EngineAdapter API used for data access in ReaxKit.** The EngineAdapter abstract base class defines a shared interface for engine detection, required-file resolution, typed data loading, and engine-appropriate writing. The *detect* method is abstract and must be implemented by each engine-specific adapter, while *load* and *write* act as dispatcher methods that route canonical ReaxKit domain data types or objects to engine-specific loader and writer implementations.

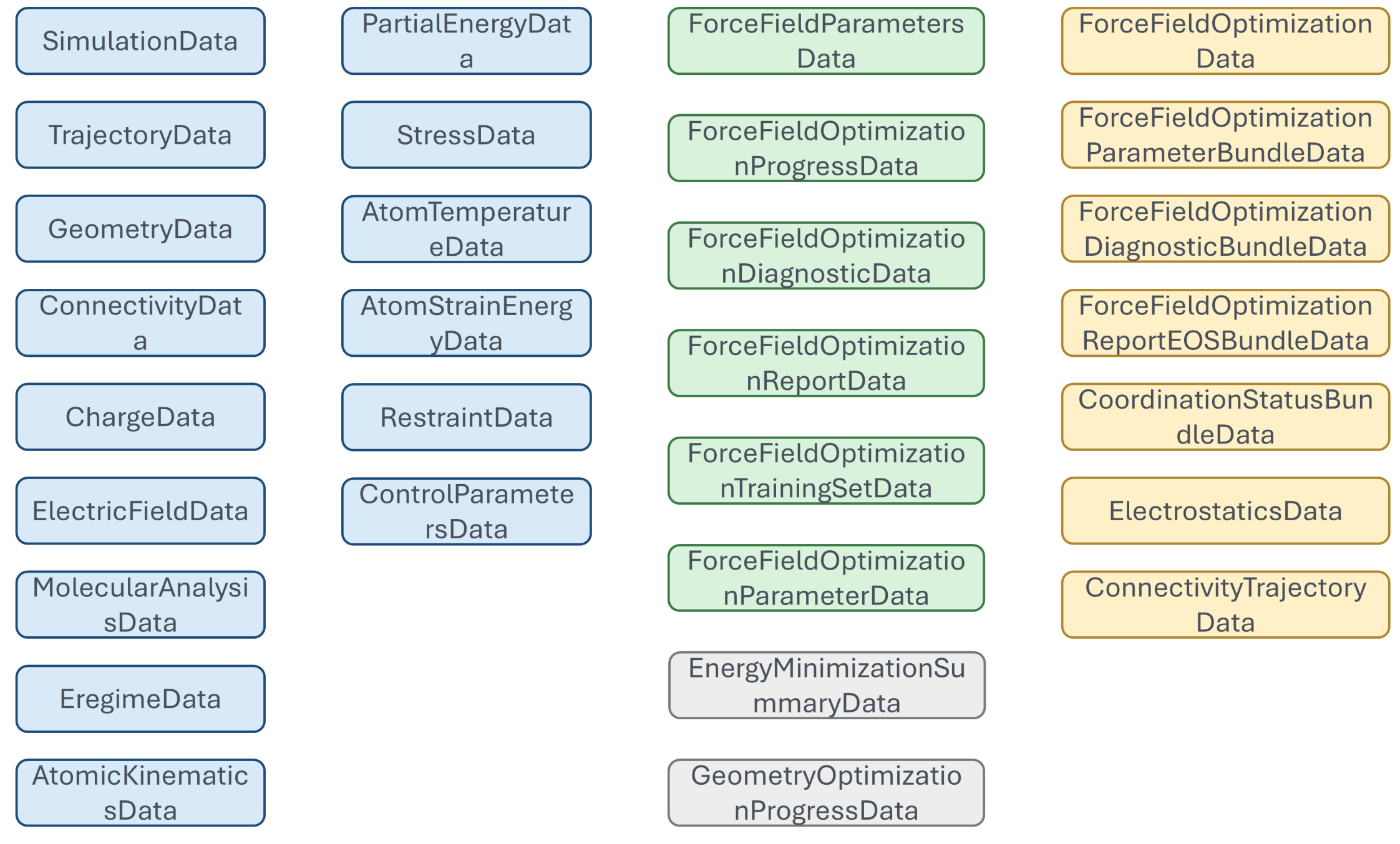

**Figure S2: Canonical domain data classes in ReaxKit.** ReaxKit represents normalized simulation information using a set of typed domain data classes. These classes provide the engine-independent data layer used by adapters, workflows, and analysis tasks, allowing outputs from different ReaxFF engines to be converted into consistent internal representations before analysis.

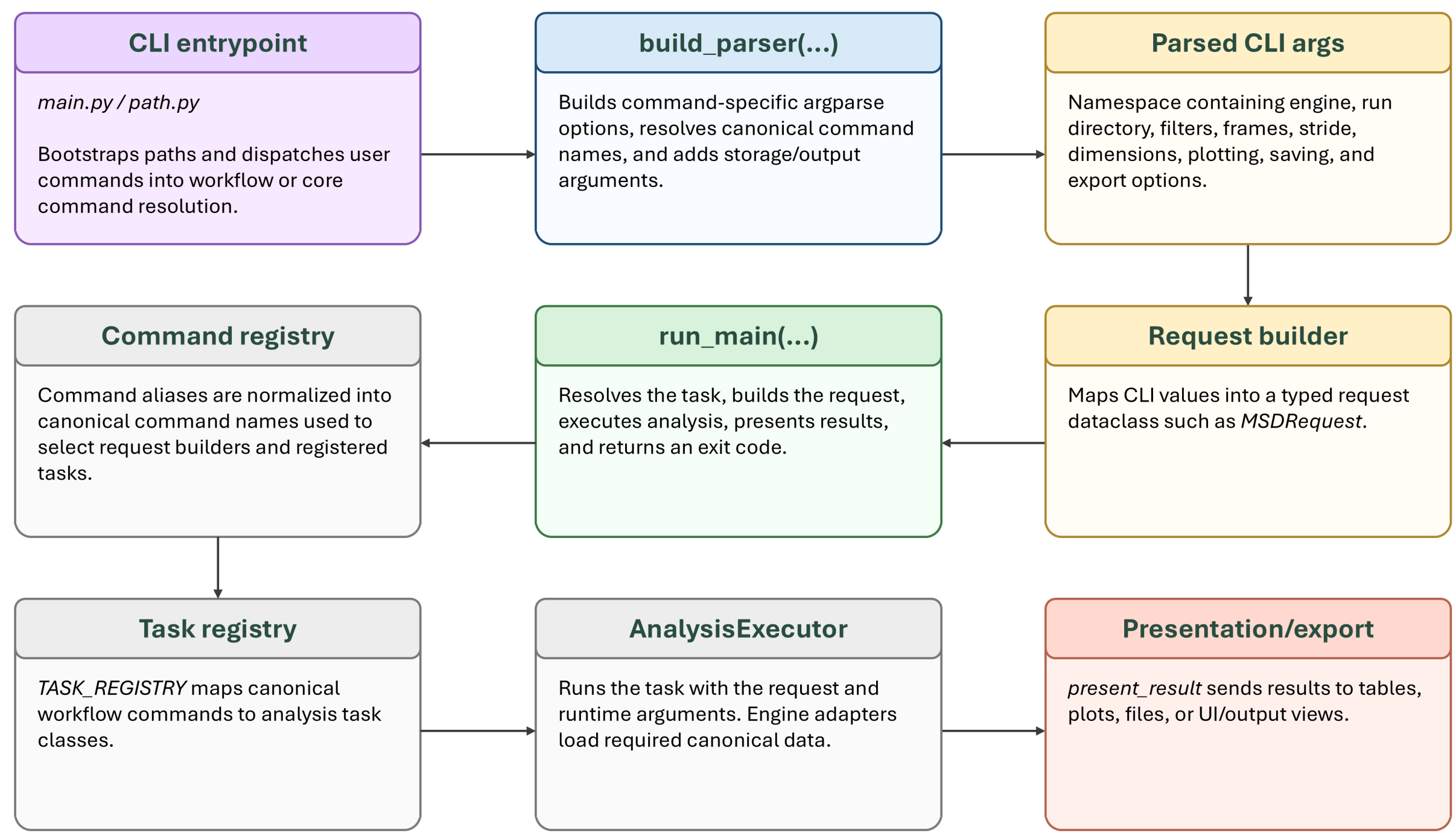

**Figure S3: Workflow automation and CLI-driven execution in ReaxKit.** Workflow modules translate user commands by configuring command-specific parsers, normalizing aliases, building typed request objects, executing registered tasks through the *analysis executor*, and dispatching results to presentation or export backends. This design keeps workflow modules focused on orchestration while leaving analysis algorithms to analyzer modules and file parsing to engine adapters.

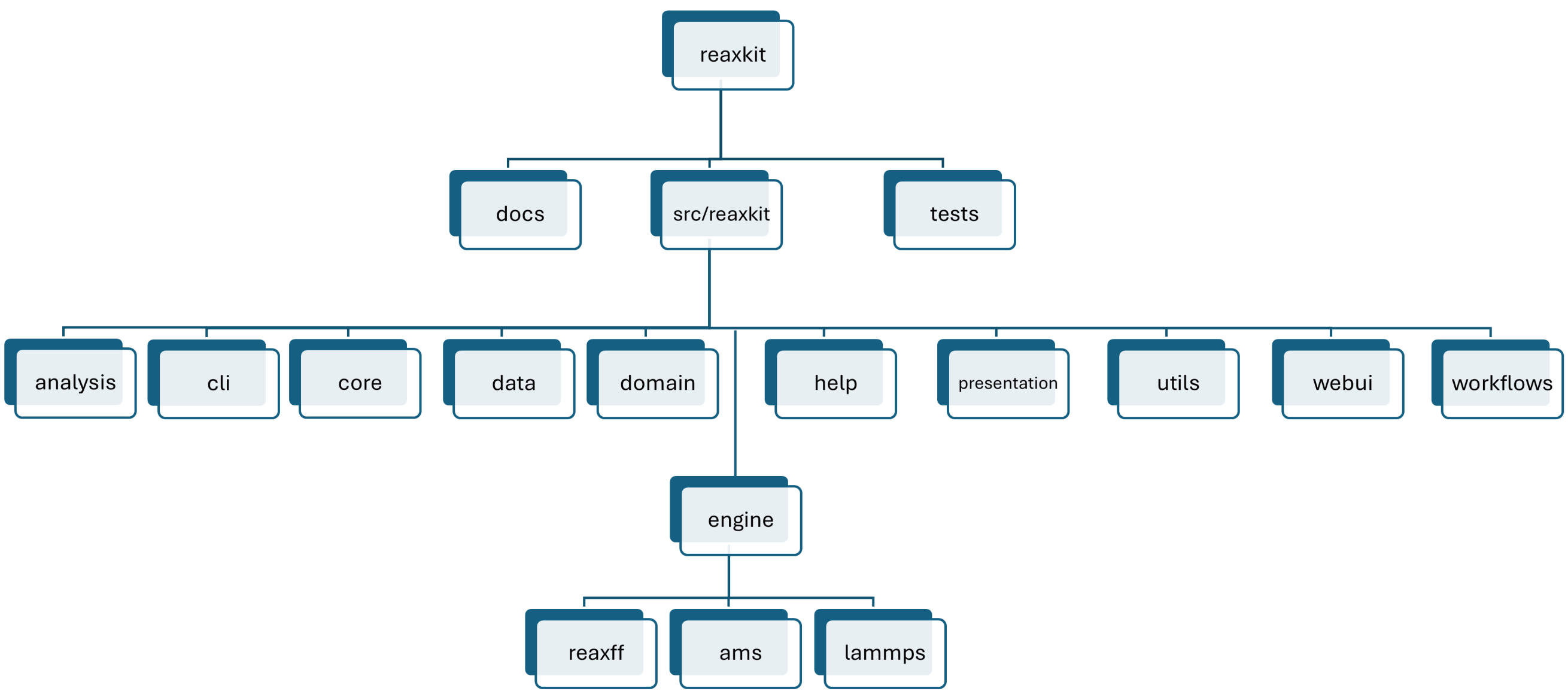

**Figure S4: Package organization and supporting components of ReaxKit.** The reaxkit repository separates documentation, source code, and tests at the top level. Within *src/reaxkit/*, the package is organized into scientific analysis modules, workflows, and several supporting components that enable usability, execution, extensibility, and maintenance. This organization separates user interaction, execution control, scientific data representation, engine-

dependent file handling, and output presentation, allowing the main analysis and workflow modules to operate through consistent internal interfaces.

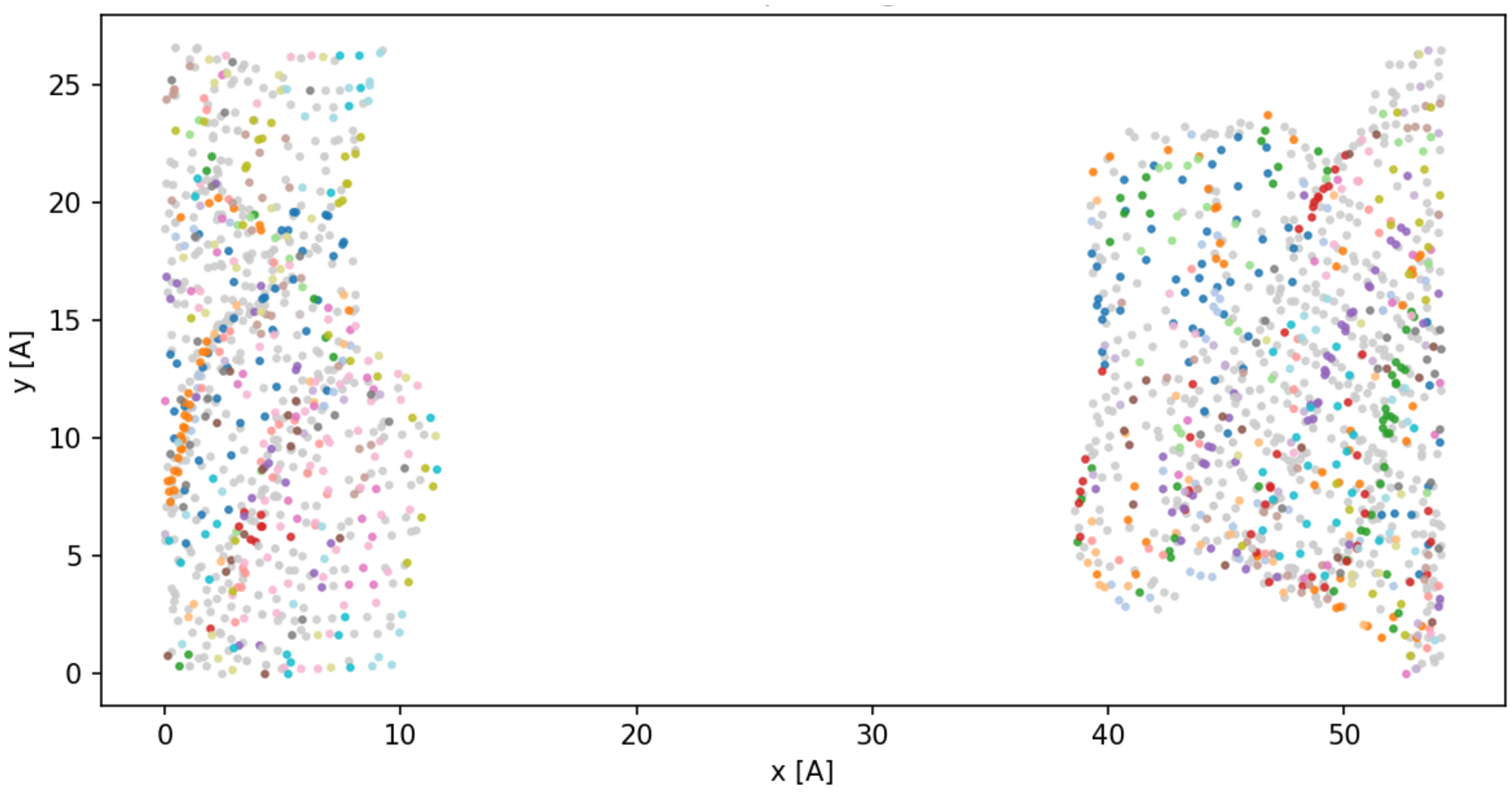

**Figure S5: Carbon orientation-domain map from local bond-orientational order.** The colors indicate local orientation-domain identifiers assigned within the carbon network. The map describes orientational fragmentation of the reconstructed carbon structure; it is not a direct measurement of La or Lc and does not assign domain identities to gas-phase atoms.

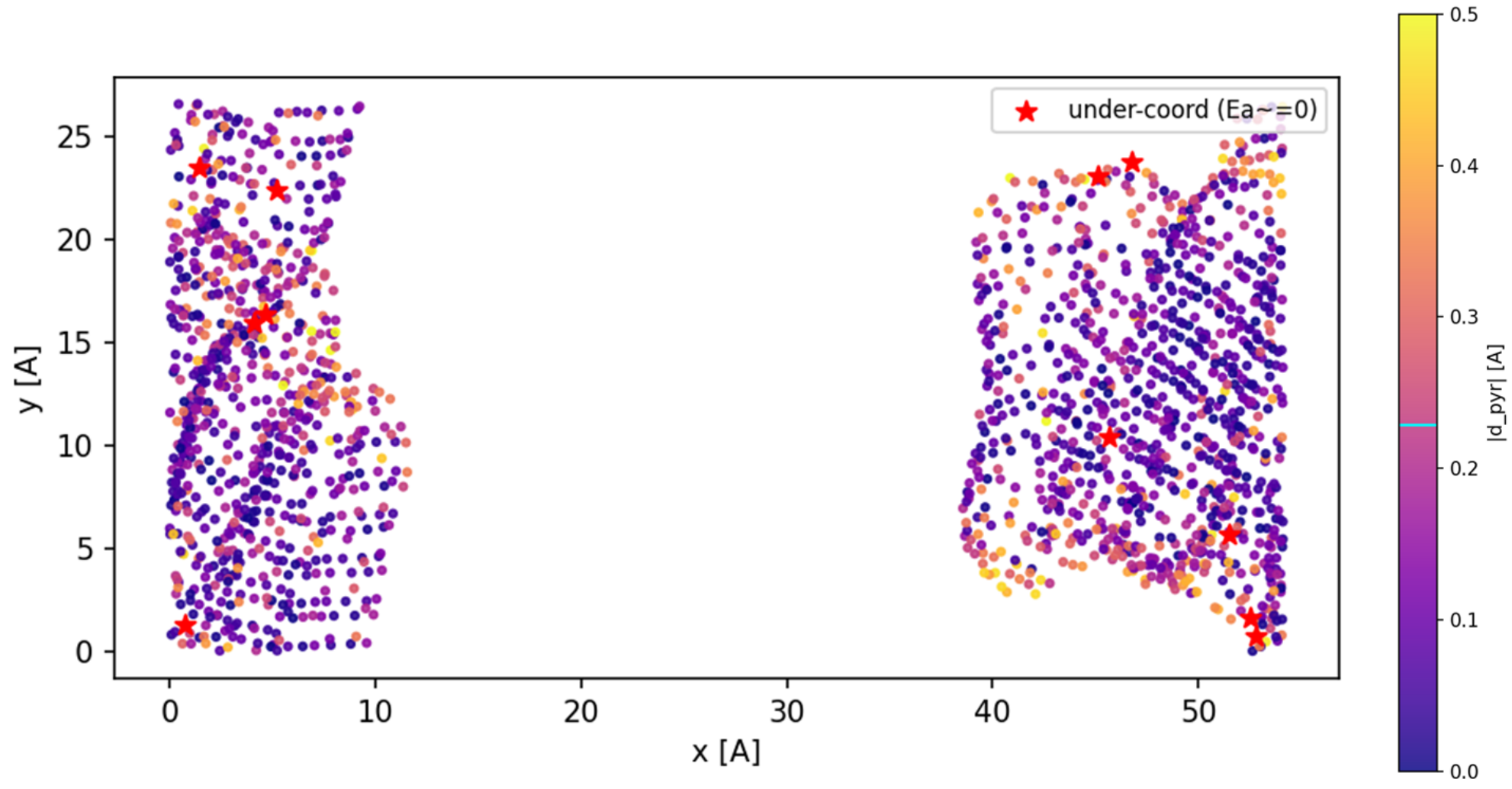

**Figure S6: Carbon pyramidalization map.** Points are colored by |d_pyr|, the absolute out-of-plane displacement of a carbon atom relative to its bonded carbon neighbors. Red stars mark under-coordinated carbon atoms. Gas-phase atoms are excluded because pyramidalization is a network descriptor.

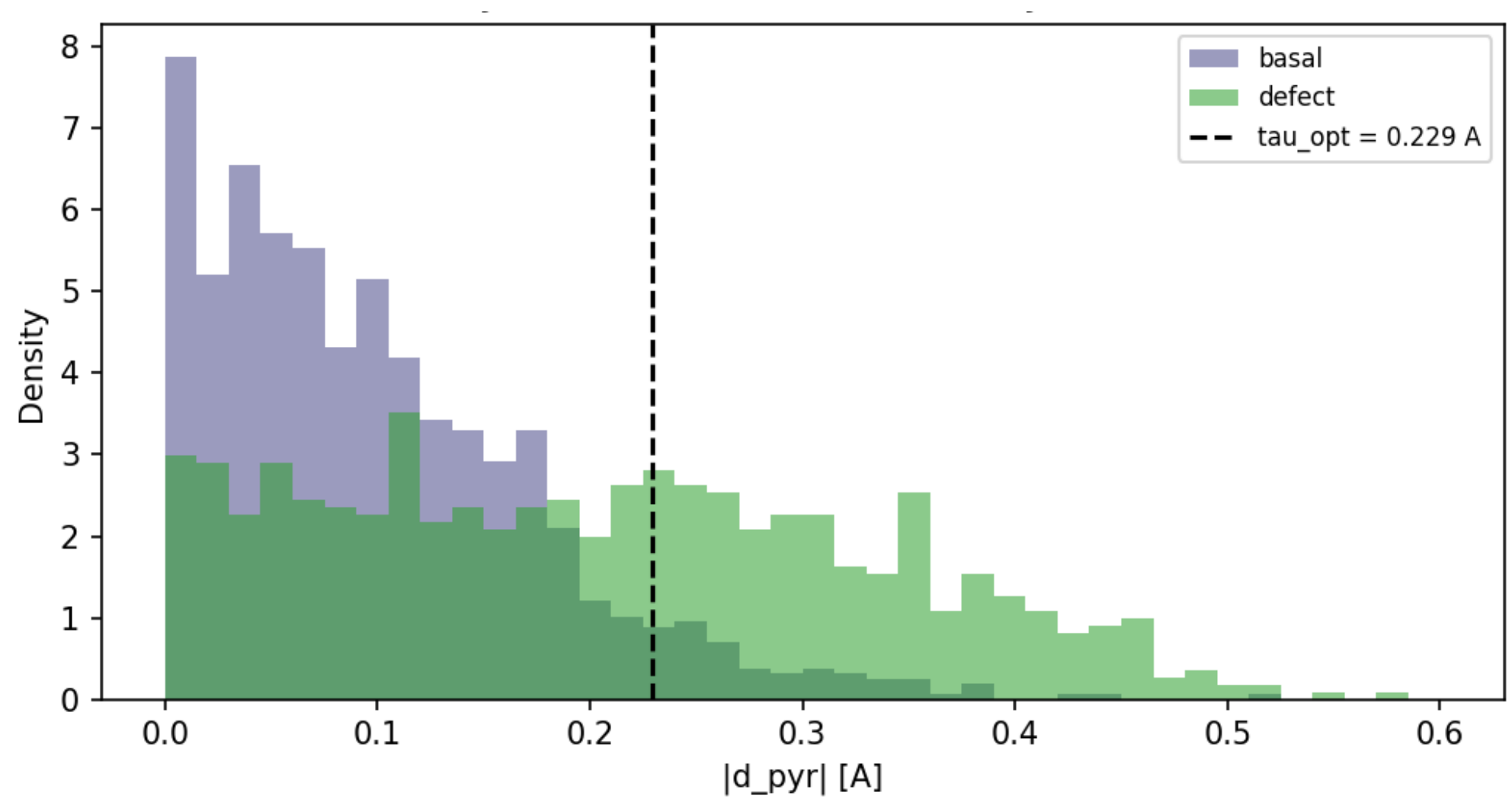

**Figure S7: Pyramidalization distribution for basal and defect-associated carbon atoms.** The vertical dashed line marks the 0.229 A reference value used in this example. Defect-associated atoms have a broader high-|d_pyr| tail than basal atoms, indicating stronger local distortion in reconstructed regions.

```yaml
study_name: mg_temp_sweep
template: "./template"
parameters:
  mg_percent: [40, 60]      # Mg percent
  temperature: [300, 500]    # Temperature (K)
replicates: 2

run:
  - stage: MM
    steps:
      - "python hybrid_generator.py Z {mg_percent}"
      - "reaxkit xtob --file xmol_hybrid_sortby_Z.xyz --dims 11.37,13.12,450 --angles 90,90,90 --sort z --output geo --copy-to-dot"
      - "sbatch submit_and_wait.sh"
    produces:
      final_geometry: "./fort.90"

  - stage: NPT
    consumes:
      initial_geometry:
        from: MM.final_geometry
        to: ./geo
    steps:
      - "reaxkit write-control --parameter mdtemp --value {temperature} --output control --copy-to-dot"
      - "sbatch submit_and_wait.sh"
      - "python fix_charges.py 32"
    produces:

  - stage: NVT
    consumes:
    steps:
      - "reaxkit write-control --parameter mdtemp --value {temperature} --output control --copy-to-dot"
      - "sbatch submit_and_wait.sh"
    produces:

analysis:
  - title: msd_atom1
    run_stage: NVT
    steps:
      - "reaxkit msd --atom-ids 1 --export results.csv"
    variables:
      iter:
        directory: "reaxkit_workspace/analysis/msd"
        file: "results.csv"
        column: "iter"
      msd:
        directory: "reaxkit_workspace/analysis/msd"
        file: "results.csv"
        column: "msd"

aggregate:
  - title: msd_atom1_aggregation
    analysis_title: msd_atom1
    x: iter
    y: [msd]
    reducer: identity
    stats: [mean, std, min, max, sem, n]
    on_missing: skip

```

**Figure S8: Example YAML configuration for a ReaxKit study-design workflow.** The YAML file defines a parameter sweep over magnesium concentration and temperature with two replicates per case, followed by a three-stage MM → NPT → NVT simulation sequence. Each stage lists the commands to execute and the artifacts produced or consumed between stages. The lower sections define post-simulation analysis, variable extraction, and aggregation rules used to summarize results across all cases and replicates.